\documentclass[a4paper, amsfonts, amssymb, amsmath, reprint, showkeys, nofootinbib]{revtex4-1}
\usepackage{amsmath,amssymb}
\usepackage{graphicx,float}
\usepackage{times,txfonts}
\usepackage{textcomp}
\usepackage{color}
\usepackage{multirow}
\usepackage{color}
\usepackage[dvipsnames]{xcolor}
\usepackage{tcolorbox}
\usepackage{soul}
\usepackage[hidelinks]{hyperref}
\usepackage{natbib}
\usepackage{nicefrac}
\usepackage{multirow}
\usepackage{blindtext}
\usepackage{changepage}
\usepackage{comment}
\usepackage[export]{adjustbox}
\usepackage[normalem]{ulem}

\newcommand{\nocontentsline}[3]{}
\let\origcontentsline\addcontentsline
\newcommand\stoptoc{\let\addcontentsline\nocontentsline}
\newcommand\resumetoc{\let\addcontentsline\origcontentsline}

\footnotetext{These authors contributed equally to this work.}

\usepackage{etoolbox}
\makeatletter
\renewcommand{\bibinfo}[2]{%
  \ifstrequal{#1}{journal}{\textit{#2}}{#2}%
}
\makeatother

\makeatletter
\newcommand{\beginsupplement}{%
  \onecolumngrid
  \@ifundefined{resumetoc}{}{\resumetoc}%

  \setcounter{section}{0}\setcounter{figure}{0}\setcounter{table}{0}\setcounter{equation}{0}%

  \renewcommand\thesection{Supplementary Note~\arabic{section}}%
  \renewcommand\thefigure{\arabic{figure}}%
  \renewcommand\thetable{\arabic{table}}%
  \renewcommand\theequation{\arabic{equation}}%
  \renewcommand{\figurename}{\textbf{Supplementary Figure}}%
  \renewcommand{\tablename}{\textbf{Supplementary Table}}%

  \def\@seccntformat##1{\csname the##1\endcsname:\quad}%

  \@ifundefined{MakeTextUppercase}{}{%
    \let\REV@MakeTextUppercase\MakeTextUppercase
    \let\MakeTextUppercase\@firstofone
  }%
  \let\REV@uppercase\uppercase
  \def\uppercase##1{##1}%

  \let\REV@numberline\numberline
  \let\REV@l@section\l@section
  \@ifundefined{l@subsection}{}{\let\REV@l@subsection\l@subsection}%
  \@ifundefined{l@subsubsection}{}{\let\REV@l@subsubsection\l@subsubsection}%

  \def\numberline##1{}%

  \providecommand*\SuppTocSectionSep{.35\baselineskip}

  \def\l@section##1##2{%
    \addpenalty\@secpenalty
    \addvspace{\SuppTocSectionSep}%
    \@dottedtocline{1}{0em}{\z@}{##1}{##2}%
  }%
  \@ifundefined{l@subsection}{}{\def\l@subsection##1##2{\@dottedtocline{2}{1.5em}{\z@}{##1}{##2}}}%
  \@ifundefined{l@subsubsection}{}{\def\l@subsubsection##1##2{\@dottedtocline{3}{3.0em}{\z@}{##1}{##2}}}%
}
\makeatother

\begin{document}

\title{Spectral tuning and nanoscale localization of single color centers in silicon via controllable strain}

\author{Alessandro~Buzzi$^{1,\dagger,*}$,  Camille~Papon$^{1,\dagger,*}$, Matteo~Pirro$^{2,3}$, Odiel~Hooybergs$^{1}$, Hamza~Raniwala$^{1}$, Valeria~Saggio$^{1}$, Carlos~Errando-Herranz$^{2,3,*}$, and Dirk~Englund$^{1,}$}
\email{Corresponding authors: abuzzi@mit.edu, cpapon@mit.edu, \\c.errandoherranz@tudelft.nl, englund@mit.edu}

\address{\vspace{0.2cm} $^1$Research Laboratory of Electronics, Massachusetts Institute of Technology, Cambridge, Massachusetts 02139, USA\\
$^2$QuTech and Kavli Institute of Nanoscience, Delft University of Technology, Delft 2628 CJ, Netherlands\\
$^3$Department of Quantum and Computer Engineering, Delft University of Technology, Delft 2628 CJ, Netherlands} 

\begin{abstract}

\noindent The development of color centers in silicon enables scalable quantum technologies by combining telecom-wavelength emission and compatibility with mature silicon fabrication. However, large-scale integration requires precise control of each emitter's optical transition to generate indistinguishable photons for quantum networking. Here, we demonstrate a foundry-fabricated photonic integrated circuit (PIC) combining suspended silicon waveguides with a microelectromechanical (MEMS) cantilever to apply local strain and spectrally tune individual G-centers. Applying up to $35$~V between the cantilever and the substrate induces a reversible wavelength shift of the zero-phonon line exceeding $100$~pm, with no loss in brightness.
Moreover, by modeling the strain-induced shifts with a digital twin physical model, we achieve vertical localization of color centers with sub-$3$~nm vertical resolution, directly correlating their spatial position, dipole orientation, and spectral behavior.
This method enables on-demand, low-power control of emission spectrum and nanoscale localization of color centers, advancing quantum networks on a foundry-compatible platform.

\end{abstract}

\maketitle

\stoptoc
 
\noindent Color centers are solid-state spin-photon interfaces that serve as core components for quantum information processing~\cite{aharonovich_solid-state_2016,awschalom_quantum_2018}.
Their long spin and optical coherence times, combined with the ability to be optically initialized and read out, make them suitable for quantum communication~\cite{bhaskar_experimental_2020,pompili_realization_2021,knaut_entanglement_2024}, simulation~\cite{randall_many-bodylocalized_2021}, and computing applications~\cite{choi_percolation-based_2019}. Silicon photonics as a host platform for color centers, such as G-centers~\cite{redjem_single_2020} and T-centers~\cite{higginbottom_optical_2022}, offers a scalable solution for quantum technologies at telecom wavelengths, driven by advanced fabrication techniques and seamless integration with photonic active components and CMOS electronics~\cite{atabaki_integrating_2018}. 
Integration of these color centers into photonic structures, such as cavities~\cite{saggio_cavity-enhanced_2024,redjem_all-silicon_2023,islam_cavity-enhanced_2023,johnston_cavity-coupled_2024} and waveguides~\cite{lee_high-efficiency_2023,prabhu_individually_2023,komza_indistinguishable_2024}, has demonstrated their potential as a scalable quantum photonic platform~\cite{simmons_scalable_2024}.

The generation of large-scale entangled states is fundamental to the advancement of quantum platforms~\cite{inc_distributed_2024, li_heterogeneous_2024}. Realizing such states within solid-state quantum emitters requires the reliable generation of indistinguishable photons. The scale and fidelity of the entangled state ultimately depend on the ability to discriminate individual color centers and control their emission wavelength to overcome the sensitivity of their properties from the local strain and charge environment~\cite{li_heterogeneous_2024, papon_independent_2023,chu_independent_2023, larocque_tunable_2024}. Recent studies have investigated the influence of strain~\cite{ristori_strain_nodate, durand_hopping_2024}, electric fields~\cite{day_electrical_2024, clear_optical_2024}, and laser irradiation~\cite{prabhu_individually_2023} on color centers in silicon, uncovering fundamental properties and viable spectral tuning mechanisms. In particular, strain tuning has emerged as a highly effective approach for controllably shifting the optical transition energies of diamond color centers, enabling reversible wavelength alignment between otherwise non‐identical emitters~\cite{meesala_strain_2018, sohn_controlling_2018, machielse_quantum_2019, brevoord_large-range_2025}.
Despite these advances, the inability to independently tune the emission spectrum of individual color centers in silicon and discriminate them based on their tuning behaviors limits the scalability and performance of the platform.

Here, we address these challenges by demonstrating reversible spectral tuning and nanoscale localization of individual G-centers in photonic circuits through local strain control. Using a suspended waveguide cantilever actuated via Micro-Electro-Mechanical System (MEMS) mechanisms, we observe hysteresis-free tuning of emission wavelengths via low-power strain control. Additionally, by modeling the spectral tuning through the piezospectroscopic model, the electromechanical behavior of the device, and the emitter's collection efficiency, we extract the color center's orientation and vertical position with nanometric resolution.

\begin{figure*}[t!]
	\centering
	\includegraphics[width=\linewidth]{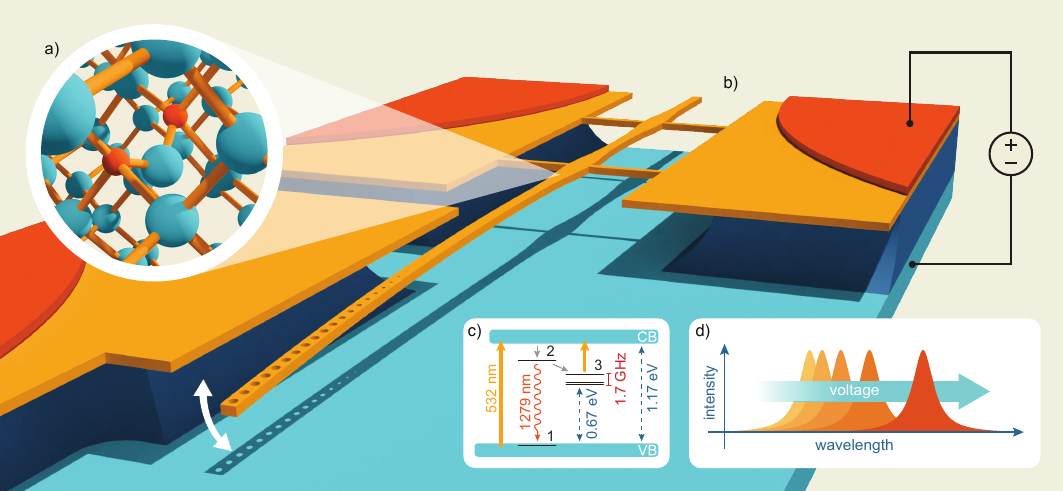}
	\caption{\textbf{Illustration of the MEMS cantilever for strain-tuning of color centers.}
    \textbf{a)}~Ball-and-stick model of the G-center, a color center in silicon, comprising two substitutional carbon atoms (orange) and an interstitial silicon atom (light blue) within the silicon crystal lattice.
    \textbf{b)}~Illustration of the suspended cantilever waveguide, held in place by tethers that also provide electrical contact, and terminated by a Bragg reflector. The waveguide can be bent through capacitive actuation by applying a voltage difference between the waveguide and the silicon handle ground plane. This controlled bending enables accurate manipulation of strain within the cantilever.
    \textbf{c)} Schematic of the silicon conduction‑band minimum (CB) and valence‑band maximum (VB), overlaid with the G‑center energy levels. 1. denotes the electronic ground state, 2. the optically excited state, and 3. the three sub‑levels of the metastable triplet ($S=1$). The zero-field splitting of the triplet is derived from Ref.~\cite{odonnell_origin_1983}, while its energy from the ground state is obtained from Ref.~\cite{udvarhelyi_identification_2021}. The diagram shows the above-band excitation and the radiative and non-radiative transitions of the color center, which are influenced by strain introduced by the cantilever actuation.
    \textbf{d)}~Conceptual plot of the G-center emission spectrum, displaying intensity as a function of wavelength. The plot shows how strain shifts the emission spectrum. The magnitude of this shift is influenced by the emitter's orientation and position within and along the cantilever.}
    \label{fig:fig1}
\end{figure*}

\begin{figure*}[]
	\centering
	\includegraphics[width=\linewidth]{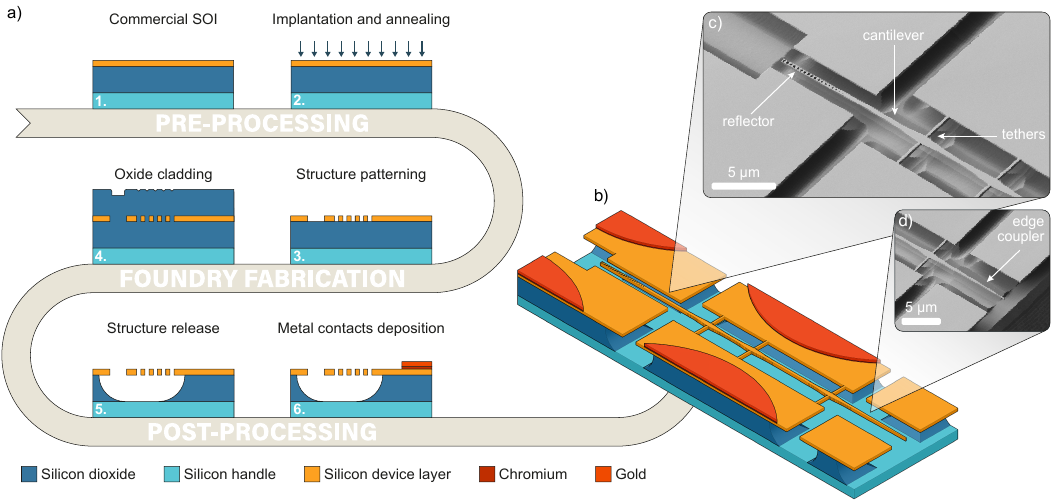}
    \caption{\textbf{Device fabrication process and micrographs.} \textbf{a)} Fabrication starts with a commercial silicon-on-insulator (SOI) chip (1). The chip undergoes carbon ion implantation followed by annealing to restore the crystalline structure and form G-centers (2). The chip is then sent to a photonics foundry for device layer etching (3) and oxide cladding deposition (4). In the final post-processing steps, MEMS structures are released via undercutting in wet etching, followed by critical point drying (5). Electrical contacts are deposited by metal evaporation through a shadow mask aligned with the pad area (6). \textbf{b)} Isometric schematic of the fabricated device, showing the cantilever terminated by a Bragg reflector at the top and an edge coupler at the bottom. The central part of the device, illustrating three of the actual ten tethers, has been shortened in this view for clarity \textbf{c)} Micrograph of the suspended MEMS cantilever. \textbf{d)} Micrograph of the linear inverse taper edge coupler. The micrographs were acquired from an equivalent device; the actual strain-tuning device was not imaged via scanning electron microscopy to avoid structural collapse due to stray charging.}
    \label{fig:fig2}
\end{figure*}

\section{Results}
\subsection*{Device concept and design}

\noindent The device's working principle, illustrated in Fig.~\ref{fig:fig1}, involves controlling strain within a suspended silicon waveguide to modulate the emission wavelength of embedded G-centers (Fig.~\ref{fig:fig1}a). The MEMS consists of a mechanical cantilever capacitively actuated by a voltage difference between the silicon device layer and the silicon handle (Fig.~\ref{fig:fig1}b). The applied voltage induces strain along the waveguide, which affects the energy levels of the G-centers (Fig.~\ref{fig:fig1}c), resulting in a shift in their emission wavelength (Fig.~\ref{fig:fig1}d).

An above-bandgap laser excites the G-centers via a confocal microscope focused on the cantilever. Emission from the color centers is collected through the waveguide, which is suspended by mechanical tethers. A photonic Bragg reflector at the end of the waveguide reflects the emission along the cantilever, increasing the collection efficiency. A linear inverse taper then couples the emitted light out, which is collected by an ultra-high numerical aperture (UHNA) fiber and detected by a superconducting nanowire single-photon detector (SNSPD). The photonic components were optimized for the fundamental quasi-TE waveguide mode using finite-difference time-domain (FDTD) simulations. Further details on the design and experimental setup can be found in Supplementary Notes 1 and 2.

\subsection*{Foundry-based fabrication}

\noindent The fabrication process consists of three main phases: preprocessing, commercial fabrication, and post-processing.

During preprocessing, we start with a $100$-oriented silicon-on-insulator (SOI) wafer. The SOI device layer is $220$~nm thick, and the bottom oxide (BOX) is $2$~µm thick. Carbon ion implantation into the device layer is followed by rapid thermal annealing (RTA) to form G-centers.

In the commercial fabrication phase, the wafer is sent to a photonics foundry for electron beam lithography patterning and reactive ion etching (RIE) of the device layer. A $2$~µm silicon dioxide cladding is deposited via plasma-enhanced chemical vapor deposition (PECVD).

Post-processing involves releasing the MEMS structures by wet etching in hydrofluoric acid (HF), removing the oxide cladding, and undercutting the structures. Critical point drying (CPD) is used to prevent collapse during drying. Chromium-gold electrical contact pads are patterned through electron beam evaporation with a shadow mask to avoid any liftoff or etching step that could damage the suspended structures (see Supplementary Note~3). Finally, the chip is wire bonded to a printed circuit board for electrical connection. The details of the fabrication process are provided in the Methods section.

Schematics of the fabrication process and final device are shown in Fig.~\ref{fig:fig2}a and Fig.~\ref{fig:fig2}b. Scanning electron micrographs of the key photonic components after fabrication are shown in Fig.~\ref{fig:fig2}c-d. Fig.~\ref{fig:fig2}c displays the MEMS cantilever waveguide, including the Bragg reflector and tethers. Fig.~\ref{fig:fig2}d shows the linear inverse taper edge coupler. The Bragg reflector is designed with a reflectance of $95$~\%. The tether transmission, measured experimentally, is $94$~\%, and the coupling efficiency of the edge coupler to the UHNA fiber is calculated to be above $12$~\%. Further details on the efficiency of the device's components are provided in Supplementary Note~4.

\subsection*{Spectral tuning of single G-centers}

\begin{figure*}[]
	\centering
	\includegraphics[width=\linewidth]{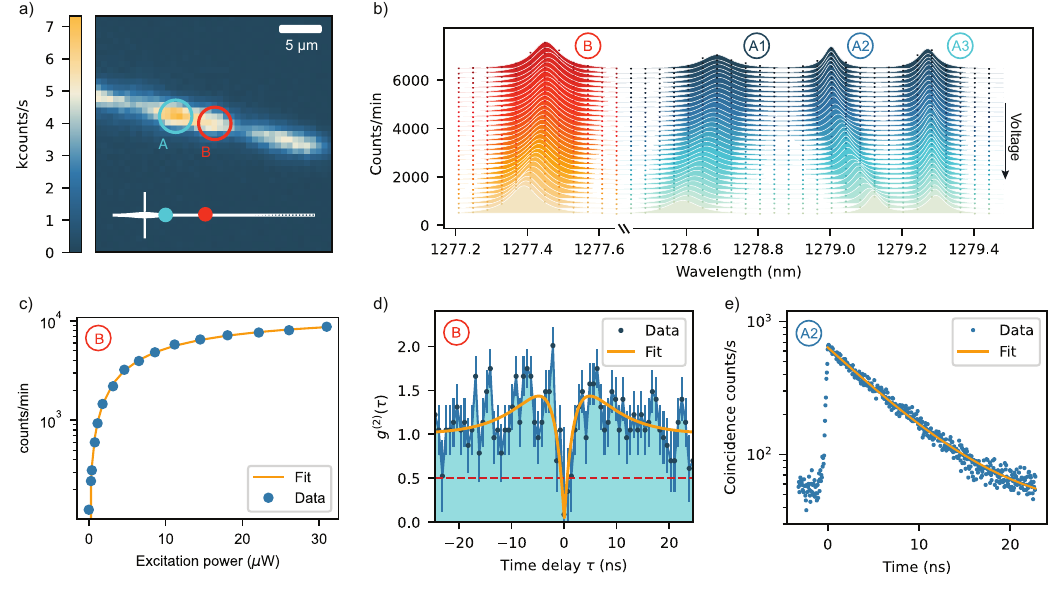}
	\caption{\textbf{Strain tuning of single waveguide-coupled G-centers.}  {\bf a)}  Photoluminescence (PL) emitted from waveguide-coupled G-centers collected from the edge coupler and detected with SNSPDs, as a function of the excitation laser position  ($\lambda_\text{exc}=532$~nm). The PL is filtered with a broadband free-space bandpass filter ($1250$-$1300$~nm). Two G-center spots within the waveguide are identified as $A$ and $B$ and are sketched in the inset. {\bf b)} Several zero-phonon lines are observed on the spectrometer at locations $A$ (blue) and $B$ (orange). The central wavelengths shift with the applied DC voltage. The white lines indicate the Lorentzian fit. {\bf c)} Integrated intensity on the spectrometer as a function of excitation power on emitter $B$. The yellow line indicates the fit to power saturation of a two-level system, from which we extract a saturation power of $P_\text{sat}=(13.6\pm0.5)$~µW. {\bf d)} Second-order correlation function of the filtered zero-phonon line shown in c), excited in continuous-wave at saturation, with a measured $g^{(2)}(0)=0.09\pm0.04$. The yellow line represents a fit to a three-level system, including a dark state, which gives rise to antibunching and bunching timescales. The dashed orange line indicates the threshold below which emission from a single emitter is demonstrated. {\bf e)} Time-resolved measurement of the zero-phonon line $A_2$ under pulsed above-band excitation, close to saturation. The lifetime extracted from a mono-exponential decay, $\tau= (6.61 \pm 0.09)$~ns, is typical of G-centers.}	
    \label{fig:Fig3}
\end{figure*}

\noindent After packaging, the device is mounted in a closed-cycle cryostat and cooled down to cryogenic temperatures ($T = 7$~K) to investigate photoemission from waveguide-coupled G-centers. An above-band continuous-wave laser ($\lambda_\text{exc}=532$~nm, see Supplementary Note 5 for additional above-band spectroscopy) is scanned around the suspended cantilever region, and emission from G-centers is collected into UHNA fibers through the tapered edge coupler. The UHNA fibers are then spliced to SMF28 fibers for further routing (see Supplementary Note 6). The color center photoluminescence (PL) is then coupled to a free-space bandpass filter ($1250$ - $1300$~nm) to suppress the excitation laser and unwanted background emission before detection on the SNSPDs (see Methods). The PL raster scan shown in Fig.~\ref{fig:Fig3}a reveals two bright locations ($A$ and $B$) in the cantilever part of the device (see inset of Fig.~\ref{fig:Fig3}a).
The unfiltered emission from the two locations is then sent to a spectrometer with a resolution of $40$~pm (see Methods) and reveals several zero-phonon lines (ZPL), as shown in Fig. \ref{fig:Fig3}b). On each excitation position, we record the spectra of the ZPLs as a function of the applied voltage between the cantilever and the substrate. As the voltage is increased from $0$~V up to $35$~V, the central wavelength of the ZPL shifts, with a sign and magnitude analyzed in the theoretical model introduced in the next section. A maximum tuning of $\delta=130$~pm is observed, with an electrical power dissipated as low as $\approx 10$ nW (see Supplementary Note 5),  and we ensure that the process is reversible by recording spectra from $35$~V back to $0$~V, as shown in Supplementary Note 5. This actuation results in a spectral tuning rate of $680$~MHz/V, and other devices investigated, as shown in Supplementary Note 7, demonstrate that rates up to $5.8$ GHz/V are achievable. This latter result, obtained by applying a strain with a combination of lateral and vertical displacement, is sufficient to bring two emitters in resonance (see Supplementary Note 7). \\
To verify the two-level system nature of the tunable ZPLs, we first investigate the saturation of the single ZPL on position $B$ by increasing the excitation laser power and recording individual spectra on the spectrometer. The integrated intensity of each peak as a function of excitation power is fitted to the saturation of a two-level system (see Methods) from which we extract a saturation power of $P_\text{sat}=(13.6\pm0.5)$~\textmu W measured before the objective, as shown in Fig.~\ref{fig:Fig3}c. Saturation curves from other emitters are provided in Supplementary Note~5, all demonstrating agreement with a model of saturation of a two-level system.

We then verify the single-photon nature of the collected PL by performing a Hanbury-Brown-Twiss experiment on the narrow-filtered ZPL (See Methods) from position $B$ at saturation. Figure~\ref{fig:Fig3}d shows the time correlation between two SNSPDs, measured after the ZPL emission is split using a 50:50 fiber beamsplitter, with a bin size of 700 ps. We measure a second-order correlation function of $g^{(2)}(0)=0.09\pm0.04$ without background subtraction and after normalization by long time delay correlation ($200$~µs), a clear signature of emission from a single emitter. Deviation from the ideal second-order correlation of a pure single-photon source $g^{(2)}(0)=0$ is attributed to residual background emission or emission from other G-centers. The error bars on each correlation, and thus on the raw measured $g^{(2)}(0)$, are given by Poissonian statistics. The raw correlations are fitted to a second-order-correlation function, which includes a bunching term~\cite{rengstl_-chip_2015} due to blinking to phenomenological dark states \cite{davanco_multiple_2014} or to the dark meta-stable triplet state identified in G-centers \cite{udvarhelyi_identification_2021}. We extract an antibunching time constant of $\tau_a= 2.7 \pm 0.5 $~ns while the bunching time constant reads $\tau_b= 6.0 \pm 0.8 $~ns. 
We measure the color center's lifetime by time-resolved measurement of PL from the ZPL excited at position $A$ with an above-band pulsed laser ($\lambda_\text{exc}=532$~nm) while filtering the $A2$ line. The result is shown in Fig.~\ref{fig:Fig3}e and is fitted to a monoexponential decay, giving a lifetime of $\tau= 6.61 \pm 0.09$~ns. A similar result, shown in Supplementary Note 5, is obtained for ZPL on position $B$, indicating a lifetime of $\tau=6.4 \pm 0.1$~ns. We highlight that previous works demonstrated that the lifetime is independent of the pulsed excitation power \cite{prabhu_individually_2023,saggio_cavity-enhanced_2024}. The lifetime value confirms that the color centers employed in this work are the genuine G-centers \cite{durand_genuine_2024,saggio_cavity-enhanced_2024}, which is guaranteed by following a similar implantation process as in Refs. \cite{prabhu_individually_2023,saggio_cavity-enhanced_2024}. Additional spectroscopy results for each investigated ZPL are available in Supplementary Note 5.

\subsection*{Model-assisted nanoscale localization}

\begin{figure*}[]
	\centering
    \includegraphics[width=\linewidth]{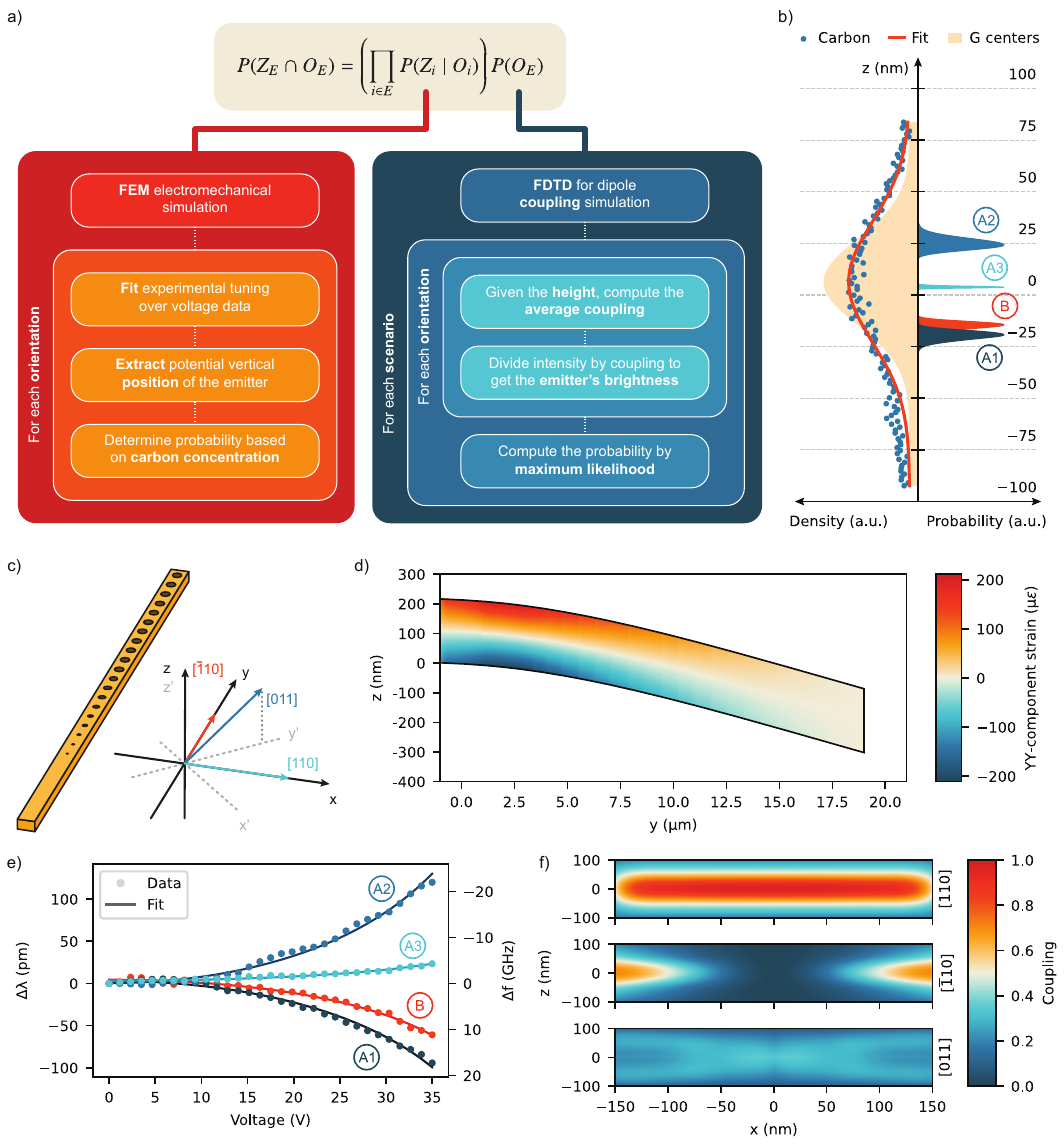}
	\caption{\textbf{Model and simulations for the vertical localization of emitters within the waveguide.}  
    \textbf{a)} Conditional probability model for estimating the vertical localization of the emitters.  
    \textbf{b)} Comparison of the normalized carbon concentration (orange line) measured via SIMS, with a peak value of \(3.3 \times 10^{18}\, \text{atoms/cm}^3\), and the derived concentration of G-centers (yellow area) within the waveguide, shown on the left, as a function of vertical position within the waveguide. The emitters' vertical localization probability distributions for the most likely scenario are depicted on the right.
    \textbf{c)} Schematic of the cantilever waveguide with the coordinate axes used in the analysis. Both the setup framework axis and the crystal framework are shown. The cantilever is oriented along the [\(\overline{1}10\)] direction.  
    \textbf{d)} Colormap showing the strain distribution along the cantilever and its displacement, extracted from FEM simulations at an applied voltage of 35~V.   
    \textbf{e)} Spectral shift for four emitters as a function of voltage applied to the cantilever. Dots represent the Lorentzian fit centers of the emission peaks, while the line shows the fitted curve derived from the FEM voltage-strain model used to determine the vertical positions of the emitters.
    \textbf{f)} Coupling to the quasi-TE fundamental waveguide mode for three different equivalent classes of emitter orientations as a function of vertical and lateral position in the waveguide.}
    \label{fig:Fig4}
\end{figure*}

\noindent The spectral response of the emitters to the mechanical actuation depends on two microscopic characteristics: the color centers' position inside the waveguide and their orientation within the crystalline lattice. The position influences the magnitude of strain applied to them, while the orientation of the defect determines how sensitive they are to the applied strain. Additionally, both characteristics affect the coupling efficiency of the emitter to the waveguide mode. By modeling these factors, we can extract the color centers' defect orientation and vertical position with nanometric resolution.

To estimate the position and orientation of the emitters, we express the joint probability conditioned on the orientations of the emitters. The variables \( Z_i \) represent the emitters' vertical positions relative to the waveguide's center, while \( O_i \) denotes the equivalence classes of their orientations. The equivalence classes of orientations, \( O_i \), group together orientations that exhibit identical behaviors for strain, such as having the same piezospectroscopic coefficients~\cite{foy_uniaxial_1981, davies_carbon-related_1983}, and dipole orientation, meaning they couple with the same efficiency into the fundamental quasi-TE waveguide mode. To collectively represent all emitters, we define \( E = \{A_1, A_2, A_3, B\} \), with \( Z_E = \{Z_i : i \in E\} \) as their positions and \( O_E = \{O_i : i \in E\} \) as their orientations. The joint probability of emitter positions and orientations is
\begin{equation}
P(Z_E \cap O_E) = P(Z_E \mid O_E)\,P(O_E),
\end{equation}
assuming the marginal independence of emitter positions given their orientations allows us to factorize the joint probability of all vertical positions. Substituting this factorization into the original expression gives  
\begin{equation}
P(Z_E \cap O_E) = \left( \prod_{i \in E} P(Z_i \mid O_i) \right) P(O_E).
\end{equation}
Here, \( P(O_E) \) represents the joint probability distribution of the emitters' orientation equivalence classes, which cannot be factorized due to the model's interdependence between the estimated emitter orientations. The intensities of the observed emitters provide information about their coupling efficiencies, which, in turn, influences the estimation of the other emitters' coupling efficiencies and dipolar orientations. 

The model used to determine each scenario’s probability is shown in Fig.~\ref{fig:Fig4}a, with the probability distribution of the emitters' vertical position presented in the right plot of Fig.~\ref{fig:Fig4}b. We estimate the probability by evaluating the likelihood of all emitter orientation and vertical position combinations, starting with the product of conditional probabilities for each emitter's position given its dipole orientation.
The coordinate system used for the model is illustrated in Fig.~\ref{fig:Fig4}c. The origin of the coordinate system is defined at the center of the waveguide cross-section for $x$ and $z$ and at the center of the last tether along the cantilever's $y$-axis. A three-dimensional finite element method (FEM) simulation calculates the strain versus voltage along the cantilever, with the result at $35$~V displayed in Fig.~\ref{fig:Fig4}d. This simulation reveals the longitudinal strain distribution, which transitions from compressive to tensile along the vertical axis. To determine the positions of the emitters along the cantilever, we use their $y$-coordinates extracted from the PL scan in Fig.~\ref{fig:Fig3}a. Different positions along the cantilever exhibit specific strain profiles, as depicted in the heatmap of Fig.~\ref{fig:Fig4}d. By fitting the simulated strain curves, evaluated at the PL spot locations along the cantilever, to the emission wavelength shifts (Fig.~\ref{fig:Fig4}e), we determine the vertical positions of the emitters associated with each equivalence class of dipole orientations. 

We assign probabilities to the positions based on normalized carbon concentration data from secondary ion mass spectroscopy (SIMS), shown in the left plot of Fig.~\ref{fig:Fig4}b (More details in Supplementary Note 8). Since two carbon atoms and an interstitial silicon are required to form a G-center, assuming a uniform interstitial silicon distribution, the G-center density is quadratic with the carbon concentration. The shaded area in the plot represents this distribution. Consequently, we assign a probability proportional to the square of the normalized carbon concentration at each vertical position. The probabilities for the vertical positions of the color centers, \( P(Z_i \mid O_i) \), are normalized across all possible orientations for each emitter.

The second term of the probability function describes the likelihood of each possible combination of the emitters' dipole orientations, represented by the joint probability of the emitters being aligned along specific directions. To evaluate this probability, we consider the intensity of the color centers' emission and the coupling related to their dipole orientations.
The right diagram of Fig.~\ref{fig:Fig4}a illustrates the corresponding model structure. The measured intensity depends on the excitation power, emitter generation rate, and collection efficiency, which has two components: \(\eta_{\text{dip}}\), the coupling efficiency between the emitter's dipole emission and the waveguide mode, and \(\eta_{\text{coll}}\), the coupling efficiency from the waveguide mode to detection. For the four emitters analyzed, excitation power and \(\eta_{\text{coll}}\) are constant, so the main factors influencing intensity are dipole orientations and generation rates. The generation rate, as reported in Ref. \cite{prabhu_individually_2023}, is assumed to follow a normal distribution. To isolate the impact of the generation rate on the emitters' intensities, the intensity is scaled by dividing it by \(\eta_{\text{dip}}\) for each potential scenario. Coupling is averaged across the cantilever’s $x$-direction to account for uniform emitters’ distribution. \(\eta_{\text{dip}}\) is computed for all dipole orientations and vertical positions via FDTD simulations (Fig.~\ref{fig:Fig4}f and more details in Supplementary Note 9). Since the dipole coupling to the quasi-TM mode is significantly lower than to the quasi-TE mode (Supplementary Note 9), we only consider the latter one in our model. Once the relative generation rates are obtained, a maximum likelihood estimation determines the most likely Gaussian distribution for these rates. The likelihood of the four generation rate samples from each distribution is computed, yielding the overall likelihood for each orientation scenario, which corresponds to \( P(O_E) \).

We combine the two components of the model to compute the probability for each scenario, evaluating the probability \( P(Z_E \cap O_E) \) for all the possible emitter positions and orientations. The distributions, detailed in Supplementary Note 10, show that the most likely scenario ($67.0$~\%) corresponds to the case with all emitters aligned along the [$110$] or [\(\bar{1} \bar{1} 0\)] direction. Marginal probabilities for these orientations show likelihoods of $98.6$~\%, $96.0$~\%, $93.4$~\%, and $68.2$~\% for emitters \( B \), \( A_1 \), \( A_2 \), and \( A_3 \), respectively. The outcome aligns with our expectation, as the brightest emitters are typically aligned along the [$110$] direction, making them more likely to be observed during the measurements. Based on these dipole orientations, the vertical positions of the emitters are estimated (right plot of Fig.~\ref{fig:Fig4}b). Monte Carlo simulations (detailed in Supplementary Note 10) are used to estimate the error in the vertical localization, accounting for uncertainties in FEM simulation, piezospectroscopic coefficients, and \( y \)-coordinate positioning. The vertical localization estimate achieves nanometric resolution with error margins below $3$~nm.

\section{Discussion}

\noindent This work tackles key challenges in scaling silicon-based quantum technologies through precise spectral tuning and nanoscale localization of individual color centers. These advancements enable the spectral alignment of multiple color centers, as evidenced by preliminary results demonstrating the tuning of two centers (see Supplementary Note 7). Furthermore, nanoscale localization facilitates comprehensive studies of local environmental effects on emitter properties, including homogeneous and inhomogeneous spectral distributions. This methodology extends to any type of color center\cite{lukin_4h-silicon-carbide--insulator_2020,bourassa_entanglement_2020,meesala_strain_2018,sarihan_photophysics_2025}, particularly silicon-based defects such as T-centers \cite{clear_optical_2024} and T-center-like emitters \cite{xiong_computationally_2024}, which share the $C_{1h}$ symmetry group with G-centers, thus opening pathways to study and control their optical and spin properties.

Future work should focus on optimizing device design to increase the maximum achievable strain, as detailed in Supplementary Note 11, by reducing cantilever lengths to achieve quadratic strain increase while optimizing the actuation area to maintain low driving voltages. Devices capable of generating larger strains will facilitate the spectral alignment of multiple emitters within a single cantilever and across different devices. These advancements in spectral alignment, when combined with coherent emission, could significantly aid the generation of highly indistinguishable photons, facilitating quantum interference between remote centers for distributed computing \cite{inc_distributed_2024} and empowering coherent emitter coupling for quantum networks \cite{sipahigil_integrated_2016,tiranov_collective_2023}.

By applying multiple strain patterns and experimentally calibrating FEM simulations, nanoscale localization could achieve three-dimensional atomic resolution. This breakthrough would unlock real-time feedback during on-demand emitter generation with local annealing techniques \cite{hollenbach_wafer-scale_2022,jhuria_programmable_2024,gu_end--end_2025} for deterministic generation, precise studies of emitter-cavity coupling, and investigations of local interactions between color centers. Additionally, it could revolutionize quantum sensing by enabling color centers to serve as sensitive probes of local material properties, enhancing our ability to study and control quantum systems at the atomic scale.

In this work, we have demonstrated individual spectral tuning of color centers, characterizing their optical response to strain applied via a MEMS structure integrated into a silicon photonic chip. We have characterized the emitters' optical properties, demonstrating excitation of single color centers, and achieved reversible tuning of individual color centers exceeding $100$~pm. This tuning range is sufficient to span the G centers' inhomogeneous distribution, as reported in Ref. \cite{gu_end--end_2025}.
Moreover, we have shown that the tuning behavior can be used as a tool to infer the position and orientation of emitters within the waveguide with nanometric precision.

This study establishes a platform for precise control and characterization of color centers in silicon, laying the groundwork for the fundamental understanding of their properties and enabling the development of advanced quantum technologies based on silicon photonics.

\section{Methods}
\subsection{Device design}

\noindent The photonic integrated circuit is designed on a silicon-on-insulator (SOI) platform with a nominal device layer thickness of $220$~nm. The waveguide width is designed as $350$~nm, with a lateral spacing of $2.5$~µm between the waveguide and the lateral slab and a $1$~µm gap between the cantilever and the slab. The various components of the device were simulated using Finite-Difference Time-Domain (FDTD) simulations with Ansys Lumerical and Tidy3D. The structure consists of a suspended cantilever, terminated by a Bragg reflector, and a waveguide suspended by $10$ tethers spaced along a $150$~µm length. The tether spacing is defined by a uniform random variable with the center of the distribution set to the maximum value below $20$~µm that ensures equal spacing, which in this case is approximately $18$~µm. The uniform random component spans a range of $2$~µm and is introduced to limit the effects of parasitic Fabry-Perot interference caused by tether scattering. At the end of the waveguide, an inverse linear taper edge coupler is employed to collect light into a fiber. The Bragg reflector, approximately $8$~µm long, features $20$ holes, $10$ for the linear adiabatic transition from the waveguide to the reflector and $10$ for the reflector itself, with a hole pitch of $400$~nm and a hole radius of $100$~nm. The simulated reflectance of the reflector is $95$~\%. The tethers provide mechanical support to the waveguide, keeping it suspended. They are formed by a Gaussian broadening up to a maximum of $720$~nm, with a standard deviation of $1.3$~µm over a transition length of $4$~µm. The simulated transmission is $96$~\% for the quasi-TE mode, with an estimated $94$~\% transmission based on high-resolution micrographs of the fabricated tether structures, as detailed in Supplementary Note 4. The edge coupler is a suspended linear inverse taper with a length of $14$~µm and a taper tip width of $110$~nm. The simulated transmission efficiency from the waveguide to the ultra-high numerical aperture fiber (UHNA3, NA = $0.35$) and mode field diameter of $3.3$~µm is $59$~\%, with a measured transmission of approximately $12$~\% for the quasi-TE mode at 1280 nm. The cantilever is $20$~µm long, with a spacing of $2.0$~µm to the ground plane, determined by the buried oxide layer thickness. The cantilever's electromechanical behavior was simulated using the finite element method (FEM) in COMSOL, which resulted in a pull-in voltage of 42.5 V and a maximum strain at pull-in of 510 µ$\varepsilon$.

\subsection{Sample Fabrication}
\noindent The fabrication process starts with an SOI wafer with [$100$] orientation.$\,^{12}$C ions are implanted into the $220$~nm device layer using an energy of $36$~keV and a fluence of \(5 \times 10^{13}\)~ions/cm². Rapid thermal annealing (RTA) is performed at $1000$°C for $20$ seconds in a nitrogen atmosphere to heal the crystalline structure and form G-centers.

Following implantation, the wafer is sent to Applied Nanotools for foundry fabrication. Electron beam (e-beam) lithography is performed using a JEOL JBX8100FS system at $100$~kV, patterning the photonic structures. The device layer is etched using a reactive ion etching (RIE) process with SF\(_6\)-C\(_4\)F\(_8\). A $2$~µm thick silicon dioxide (SiO\(_2\)) cladding is deposited using plasma-enhanced chemical vapor deposition (PECVD) at 300°C.

Post-processing takes place in the MIT.nano cleanroom. Structures are released by etching the cladding and buried silicon dioxide in a $49$~\% HF solution for $80$ seconds, resulting in an undercut of approximately $2$~µm. The chip is transferred from HF to water and then to isopropanol (IPA) while submerged in liquid. Critical point drying (CPD) is employed using CO\(_2\) to prevent device collapse by gradually replacing the IPA with liquid CO\(_2\). The drying process from liquid CO\(_2\) prevents phase transitions that would otherwise occur during evaporation, thereby avoiding the capillary forces that could cause the suspended structures to collapse.
Electrical pads, consisting of $50$~nm chromium and $200$~nm gold, are patterned using electron beam evaporation (Temescal FC2000) through a shadow mask. The shadow mask is fabricated from a $0.1$~mm ($4$~mil) steel sheet using the LPKF ProtoLaser U4 laser cutter, with $50$~µm 
holes to define the electrical pads. The mask is aligned to the chip under a microscope for alignment with tens of micrometers precision (see Supplementary Note 3 for further details) and held approximately $100$~µm above the chip during evaporation.
The sample is finally glued with silver paste onto a PCB, and the bonding pads are contacted with an aluminum wire bonder.

\subsection{Experimental Method for Spectral Tuning and Spectroscopy}
\noindent The bonded sample is mounted in a closed-cycle cryostat (Montana Instrument S50) equipped with an optical window for excitation and fiber-feedthrough for side collection (see Supplementary Note 2). The excitation laser and light source for imaging are directed to the sample through a microscope objective with a numerical aperture of $0.55$. Galvanic mirrors are positioned close to the objective to scan the excitation laser position while conserving focus. SNSPDs (Photonspot) optimized for $1550$ nm (24\% detection efficiency at 1280 nm) are used for low-time jitter ($150$~ps) detection of single photons, combined with a high time resolution time tagger (Swabian Instruments Timetagger 20). Spectra are recorded on an Oxford Instrument spectrometer equipped with a nitrogen-cooled camera (PyLon IR CCD) with an integration time of $60$~s and 1\% detection efficiency at 1280 nm. DC voltage is applied through a voltage source (Keithley $2400$), and upward and downward sweeps are recorded. Photoluminescence emitted in the waveguide mode is collected with the UHNA3 fiber.
\subsubsection{Photoluminescence raster maps}
\noindent The PL maps presented in Fig.~\ref{fig:Fig3}a are obtained by triggering the recording of photon counts on a time tagger channel by the voltage applied on the galvanic mirrors, with an integration time of $100$~ms. In this case, the collected emission is filtered through a free-space filter ($1250$-$1300$~nm) of $60\ \%$ efficiency to remove background luminescence from the signal and then fiber-coupled to a single SNSPD.  
\subsubsection{Second-order correlation function and lifetime}
\noindent To isolate a single ZPL, the collected emission is sent to a fiber-based tunable narrowband filter (WL Photonics) with a $0.1$~nm bandwidth, combined with a wavelength demultiplexer with output at $1280$~nm for suppressing background light. The filtered ZPL is then coupled to a 50:50 fiber-based beam-splitter, whose outputs are coupled to two SNSPDs. The function used to fit the normalized correlation reads
\begin{align}
g^{(2)}(\tau) = (1-Ae^{-|\tau|/\tau_a})\cdot(1+Be^{-|\tau|/\tau_b}),
\end{align}
where $A$ and $B$ are the antibunching and bunching coefficients, respectively, associated with corresponding time constants $\tau_a$ and $\tau_b$. From a non-linear least squares fitting method, we extract an antibunching constant of $A=0.98\pm0.09$ with a fixed bunching constant of $B=1.56$.

For the time-resolved measurement, we switch the excitation laser to a pulsed broadband laser (SuperK, NKT Photonics) equipped with a tunable bandwidth filter, which we center at $532$~nm. The lifetime is then fitted to a monoexponential as
\begin{align}
    I(t)= e^{-t/\tau}+c,
\end{align}
for $t>t_0$, where $c$ accounts for the constant background and $\tau$ is the total lifetime of the G-center, containing both radiative and non-radiative decays. 
\subsubsection{Saturation of a two-level system}
Emission into the ZPL as a function of excitation power is recorded by coupling the collected photons to the spectrometer, without additional filtering. The ZPL peaks are then integrated and fitted to
\begin{align}
    I_\text{int}(P)=I_{\text{max}}\frac{P}{P+P_{\text{sat}}}
\end{align}
to extract the saturation power.

\subsection{Nanoscale Localization Model}
\subsubsection{Piezospectroscopic Model}
\noindent The G-center is characterized by two substitutional carbon atoms bonded to an interstitial silicon atom $\text{Si}_{\text{int}}$. Through experimental studies \cite{thonke_new_1981} and first-principles analysis \cite{udvarhelyi_identification_2021}, it has been confirmed that the defect exhibits monoclinic-I $C_{1h}$ symmetry and behaves as a linear $\pi$ oscillator perpendicular to the symmetry plane. The C-C bond can be oriented along four equivalent crystal directions: $[111],[\overline{1}11],[1\overline{1}1],[\overline{1}\overline{1}1]$. Additionally, the $\text{Si}_{\text{int}}$ atom can occupy six possible configurations around the C-C bond. Due to the low symmetry of this defect, the multiplicity of orientational degeneracy is $24$, from the ratio between the orders of the crystal symmetry group $O_{h}$ for silicon and of the center of the defect, representing the number of possible equivalent ordinations inside the crystal.\\ 
Under external stress, the energy transitions shift for a non-cubic point group in a cubic crystal can be expressed using the equation
\begin{align}
\Delta E = \sum_{i,j=x,y,z} A_{i,j} \sigma_{i,j}^{\text{ext}},
\end{align}
where $\Delta E$ corresponds to the energy shift for the identity orientation in the crystal coordinate system, the $A_{i,j}$ coefficients form a second-rank symmetric tensor, and $\sigma_{i,j}^\text{ext}$ represents the external stress components. Once the symmetry of the defect is established, this expression can be reformulated using the piezospectroscopic model developed for non-cubic defects in cubic crystals \cite{kaplyanskii_noncubic_1967}. In this model, a matrix $A_{p}$ is associated with the stress tensor. For a monoclinic-I defect with a plane aligned along ($110$), $A_{p}$ assumes the following shape
\begin{align}
A_p =
\begin{pmatrix}
A_2 & A_3 & -A_4 \\
A_3 & A_2 & A_4 \\
-A_4 & A_4 & A_1
\end{pmatrix}.
\end{align}
By performing the matrix product with the stress tensor, the derived final expression is 
\begin{align}
\Delta E = A_1 \sigma_{zz}^{\text{ext}} + A_2 (\sigma_{xx}^{\text{ext}} + \sigma_{yy}^{\text{ext}}) + 2A_3 \sigma_{xy}^{\text{ext}} + 2A_4 (\sigma_{yz}^{\text{ext}} - \sigma_{zx}^{\text{ext}}).
\end{align}
Applying uniaxial strain along a specific crystal direction breaks the symmetry of the defect and lifts the orientational degeneracy, resulting in an energy splitting. The number of splits depends on the direction of the applied strain. For a strain applied along the [$\overline{1}10$] direction, the nonzero components of the strain tensor in crystal coordinates are $\epsilon_{xx}$, $\epsilon_{yy}$, $\epsilon_{xy}$, with an additional $\epsilon_{zz}$ component arising from the Poisson's ratio of silicon (see Supplementary Note 12).  
Using FEM analysis, the magnitude of the applied uniaxial strain in the cantilever can be calculated, and the corresponding strain tensor in crystal coordinates can be determined. The $A_{i}$ coefficients are obtained from previous experimental studies \cite{foy_uniaxial_1981}, and the terms of the elastic matrix of silicon can be used to compute the strain-stress conversion. By applying the piezospectroscopic model and accounting for all possible defect rotations for a monoclinic-I symmetry, four distinct shift rates can be identified, depending on the orientation of the defect within the crystal (see Supplementary Note 13).

\subsubsection{Localization Error by Monte Carlo Simulation}
\noindent The error in the vertical localization of the emitters is assessed through Monte Carlo simulations and arises from three primary sources of uncertainty: the estimation of the position along the cantilever (\(y\)-direction), the accuracy of the FEM simulation of maximum strain values, and the variability in the piezospectroscopic model constants. For emitters located at spot $A$ and spot $B$ (as shown in Fig.~\ref{fig:Fig3}a), the uncertainty in the $y$-direction accounts for errors in the positioning and size of the excitation laser spot. Standard deviations of $500$~nm and $750$~nm are used for the positions of spots $A$ and $B$, respectively, with spot $B$ exhibiting higher uncertainty due to its farther distance from the tether and the resulting less precise positioning along the cantilever. These values approximately correspond to twice the size of the diffraction-limited spot. A $20$~\% uncertainty is applied to the FEM maximum strain values, based on literature estimates \cite{song_effect_2009}. The variability in the piezospectroscopic constant is quantified using the sample standard deviation of literature values~\cite{foy_uniaxial_1981, davies_carbon-related_1983}, resulting in a $4.7$~\% error. All sources of uncertainty are modeled as Gaussian noise in the Monte Carlo simulations. The results of these simulations are displayed in Supplementary Note 10.

\subsubsection{G-Centers Concentration}
\noindent To estimate the relationship between carbon concentration and G-centers' formation, we model the density of G-centers using a rate equation that accounts for the reactants involved in their formation. The G-center is formed by two substitutional carbon atoms ($C_{\text{sub}}$) and one interstitial silicon atom ($\text{Si}_{\text{int}}$). The rate equation for the concentration of G-centers, denoted as $[G]$, is given by
\begin{equation}
[G] = k \cdot [C_{\text{sub}}]^2 \cdot [\text{Si}_{\text{int}}],
\end{equation}
where $k$ is the formation efficiency, $[C_{\text{sub}}]$ is the concentration of substitutional carbon atoms, and $[\text{Si}_{\text{int}}]$ is the concentration of interstitial silicon atoms. Assuming the concentration of substitutional carbon atoms linearly proportional to the overall carbon concentration $[C]$ and the concentration of interstitial silicon to be constant within the waveguide, the G-center concentration will depend quadratically from the carbon concentration, such that \([G] \propto [C]^2\).

\subsubsection{Conditional Probability Factorization}

\noindent The term \( P(Z_E \mid O_E) \) is factorized under the assumption of marginal independence of emitter positions given their orientations. The assumption means that conditioned on their orientations, the position of one emitter is independent of the positions and orientations of the other emitters, leading to
\begin{equation}
P(Z_E \mid O_E) = \prod_{i \in E} P(Z_i \mid O_i),
\end{equation}
where \( P(Z_i \mid O_i) \) is the probability that emitter \( i \) is at position \( z_i \), given that its orientation belongs to the equivalence class \( O_i \). The factorization relies on two assumptions: (1) the positions of emitters are independent given their orientations, i.e., for \( i \neq j \), \( P(Z_i, Z_j \mid O_E) = P(Z_i \mid O_E) P(Z_j \mid O_E) \); and (2) the position of an emitter is independent of the orientations of other emitters, implying \( P(Z_i \mid O_E) = P(Z_i \mid O_i) \).

\section{Data Availability}
\noindent All datasets supporting the findings of this study, sufficient to reproduce the plots and analyses in the manuscript, are available on Zenodo under DOI: \href{https://doi.org/10.5281/zenodo.16989979}{10.5281/zenodo.16989979}.

\section{Code Availability}
\noindent The analysis code and Jupyter notebooks required to process the data and generate the figures in this study are available at \href{https://doi.org/10.5281/zenodo.16989979}{10.5281/zenodo.16989979}.

\section{Acknowledgments}
\noindent The authors acknowledge S.~Nagle for assisting with the fabrication of the holder and mask for the shadow-mask evaporation and I.~Christen for manufacturing the fiber holder for the cryostat, providing laboratory support, and contributing to valuable discussions. The authors also acknowledge S.~Gyger and H.~Larocque’s contributions to the simulations during the early-stage design. I.~Berkman is acknowledged for proofreading the manuscript and offering constructive feedback. Finally, the authors thank S.~Patom\"{a}ki for assistance with the chip wire bonding.

This work was supported by the NSF Convergence Accelerator program (Award No. 2134891). C.P. acknowledges support from the NSF
Engineering Research Center for Quantum Networks (Co-operative Agreement No. 1941583). M.P. and C.E-H. acknowledge financial support from the Dutch Research Council (Project No. NGF.1623.23.027). V.S. acknowledges support from the Air Force Office of Scientific Research (AFOSR) under Award No. GR108261. D.E. acknowledges support from the NSF RAISE TAQS program.
This material is based on research sponsored by the Air Force Research Laboratory (AFRL), under agreement number FA8750-20-2-1007. 
The U.S. Government is authorized to reproduce and distribute reprints for Governmental purposes notwithstanding any copyright notation thereon.
The views and conclusions contained herein are those of the authors and should not be interpreted as necessarily representing the official policies or endorsements, either expressed or implied, of the Air Force Research Laboratory (AFRL), or the U.S. Government.

\section{Author Contributions}
\noindent A.B. and C.E-H. conceptualized the device. A.B., C.E-H., and H.R. conducted simulations and design work, while A.B. was responsible for the chip layout. The preprocessing of fabrication was carried out by A.B. and C.E-H., with postprocessing by A.B. and C.P. Experimental characterization and data analysis were performed by C.P. and A.B., with support from O.H. and V.S. The physical model for nanoscale localization was developed by A.B., in collaboration with M.P. for the piezospectroscopic model and C.P. for the color center coupling. C.E-H., V.S., and D.E. supervised the work. All authors contributed to the writing of the manuscript.

\bibliography{bibliography}

\clearpage
\part*{}
\beginsupplement

\begin{center}
    \Large \textbf{Supplementary Notes}\\[1em]
    \large \textbf{Spectral tuning and nanoscale localization of single color centers in silicon via controllable strain}\\[1em]
    \normalsize
    Alessandro Buzzi$^{1}$, Camille Papon$^{1}$, Matteo Pirro$^{2,3}$, Odiel Hooybergs$^{1}$, Hamza Raniwala$^{1}$, Valeria Saggio$^{1}$, Carlos Errando-Herranz$^{2,3}$, and Dirk Englund$^{1}$\\[0.5em]
    $^1$\emph{Research Laboratory of Electronics, Massachusetts Institute of Technology, Cambridge, Massachusetts 02139, USA}\\
    $^2$\emph{QuTech and Kavli Institute of Nanoscience, Delft University of Technology, Delft 2628 CJ, Netherlands}\\
    $^3$\emph{Department of Quantum and Computer Engineering, Delft University of Technology, Delft 2628 CJ, Netherlands}\\
\end{center}

\tableofcontents

\section{Device design}

\begin{figure*}[bhtp]
	\centering
	\includegraphics[width=\linewidth]{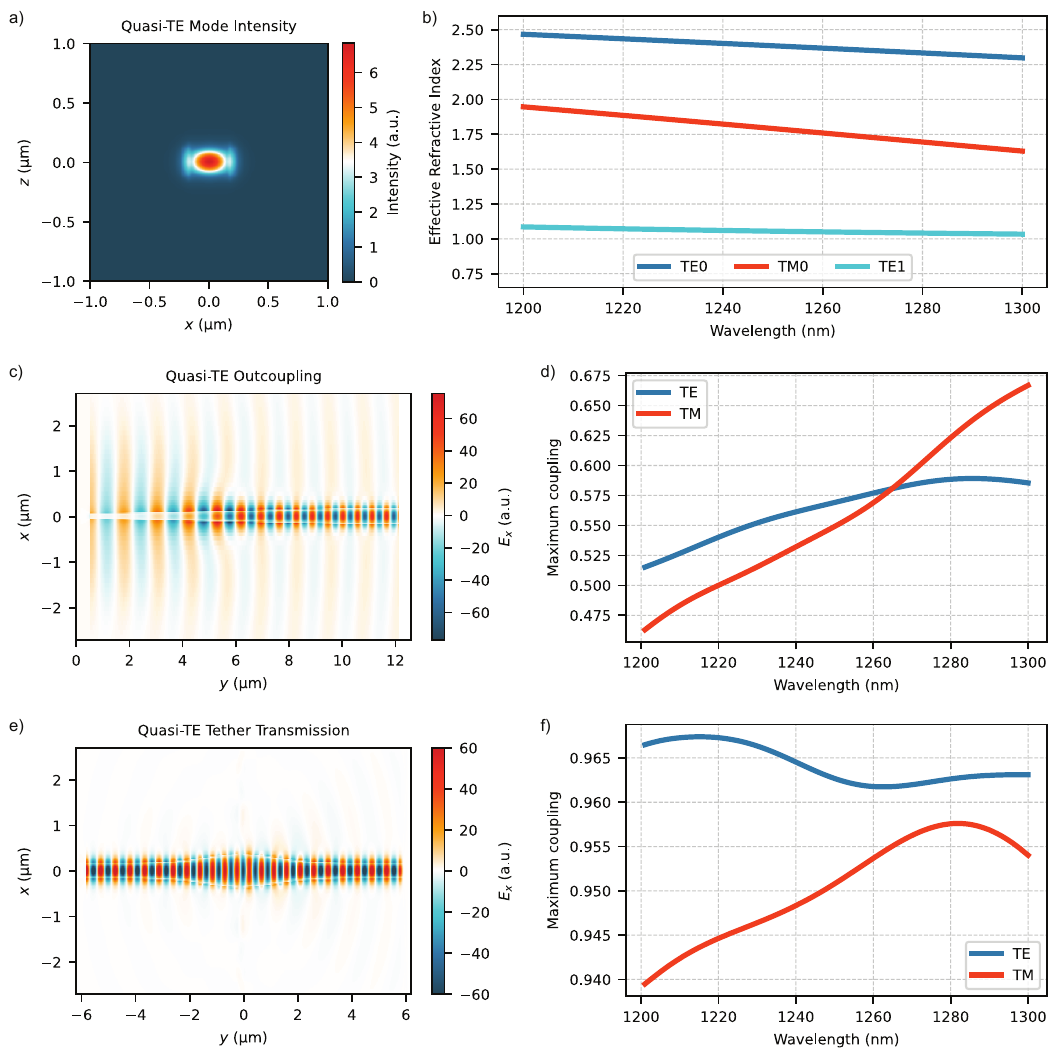}
    \caption{FDTD simulations for the design of photonic components. \textbf{a)} Mode profile intensity of the quasi-TE mode of the waveguide. \textbf{b)} Effective refractive index of waveguide modes over wavelength. \textbf{c)} Electric field $E_x$ through the inverse linear taper. \textbf{d)} Maximum transmission efficiency from a Gaussian source to waveguide modes. \textbf{e)} Electric field $E_x$ through the tether. \textbf{f)} Transmission through the tether versus wavelength for waveguide modes.}
    \label{fig:fdtd}
\end{figure*}

In this section, we present additional details about the design and simulation of the photonic components of the device. The layout of the photonic structures was created using the Python package PHIDL \cite{mccaughan_phidl_2021}, and the simulations presented here were performed with Tidy3D. The waveguide, which is $350$~nm wide and $220$~nm thick, supports a fundamental quasi-TE mode, shown in Fig.~\ref{fig:fdtd}a,  and a quasi-TM mode. At a wavelength of $1279$~nm, the effective refractive index is $2.34$ for the quasi-TE mode and $1.70$ for the quasi-TM mode, as shown in Supplementary Fig.~\ref{fig:fdtd}b.

The inverse linear taper, designed to enable efficient mode matching between the Gaussian input source and the waveguide modes, is $10$~µm long and tapers to a tip width of $110$~nm. Supplementary Fig.~\ref{fig:fdtd}c illustrates the propagation of the electric field $E_x$ through the taper. The maximum coupling efficiency through the taper at $1279$~nm is $59$~\% for the quasi-TE mode and $62$~\% for the quasi-TM mode, as shown in Supplementary Fig.~\ref{fig:fdtd}d.

The tether structure provides mechanical support while minimizing optical losses. Supplementary Fig.~\ref{fig:fdtd}e shows the electric field $E_x$ as the quasi-TE mode propagates through the tether. At a wavelength of $1279$~nm, the transmission through the tether is $96.3$~\% for the quasi-TE mode and $95.8$~\% for the quasi-TM mode, as shown in Supplementary Fig.~\ref{fig:fdtd}f.

The reflector provides reflection around $1279$~nm with low optical scattering thanks to the linear taper design, shown in Supplementary Fig.~\ref{fig:reflector}a. Supplementary Fig.~\ref{fig:reflector}c-d shows the reflection and transmission of the reflector around $1279$~nm for quasi-TE and quasi-TM modes, agreeing with the band structure of the reflector unit cell. At $1279$~nm, the reflection is about $95.35$~\% for the quasi-TE mode and about $31.5$~\% for the quasi-TM mode.

\begin{figure*}[tbhp]
    \centering
    \includegraphics[width=\linewidth]{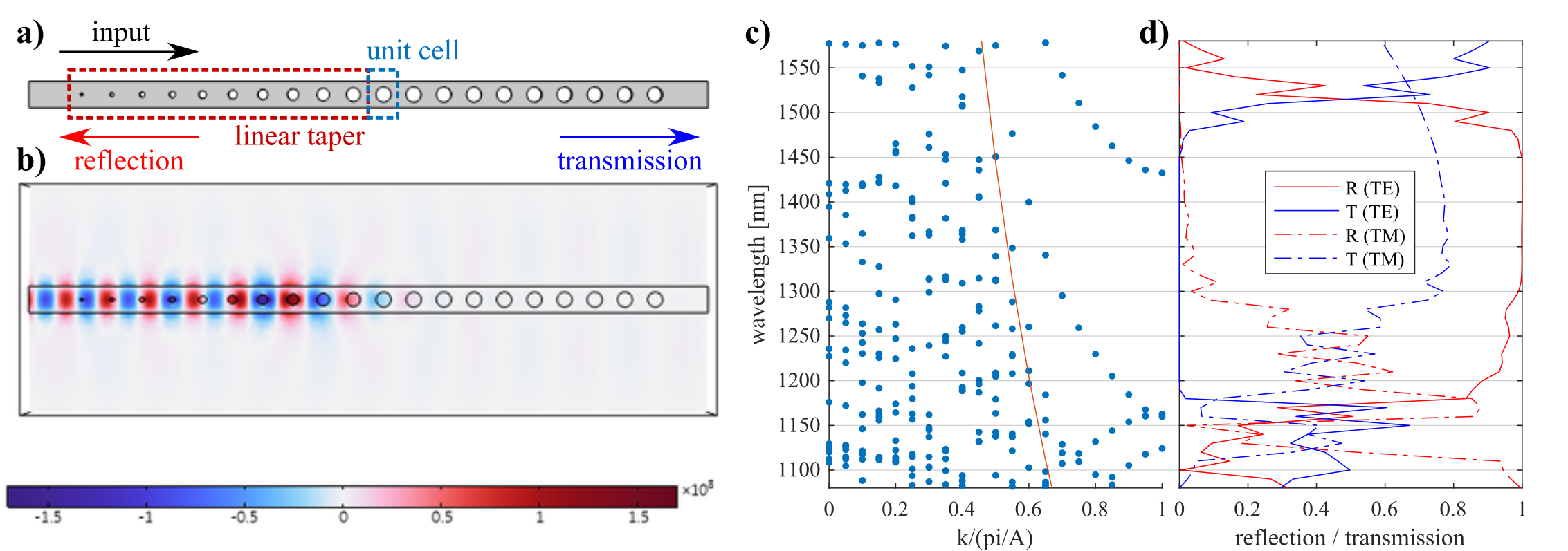}
    \caption{FEM simulation of the reflector using COMSOL Multiphysics 6.2. \textbf{a)} reflector design and \textbf{b)} associated y-component of the quasi-TE mode electric field simulation. \textbf{c)} bandstructure of the unit cell identified in a). \textbf{d)} reflection and transmission of quasi-TE and quasi-TM modes input on the reflector.}
    \label{fig:reflector}
\end{figure*}

\section{Experimental setup}

\begin{figure*}[tbhp]
	\centering
	\includegraphics[width=17cm]{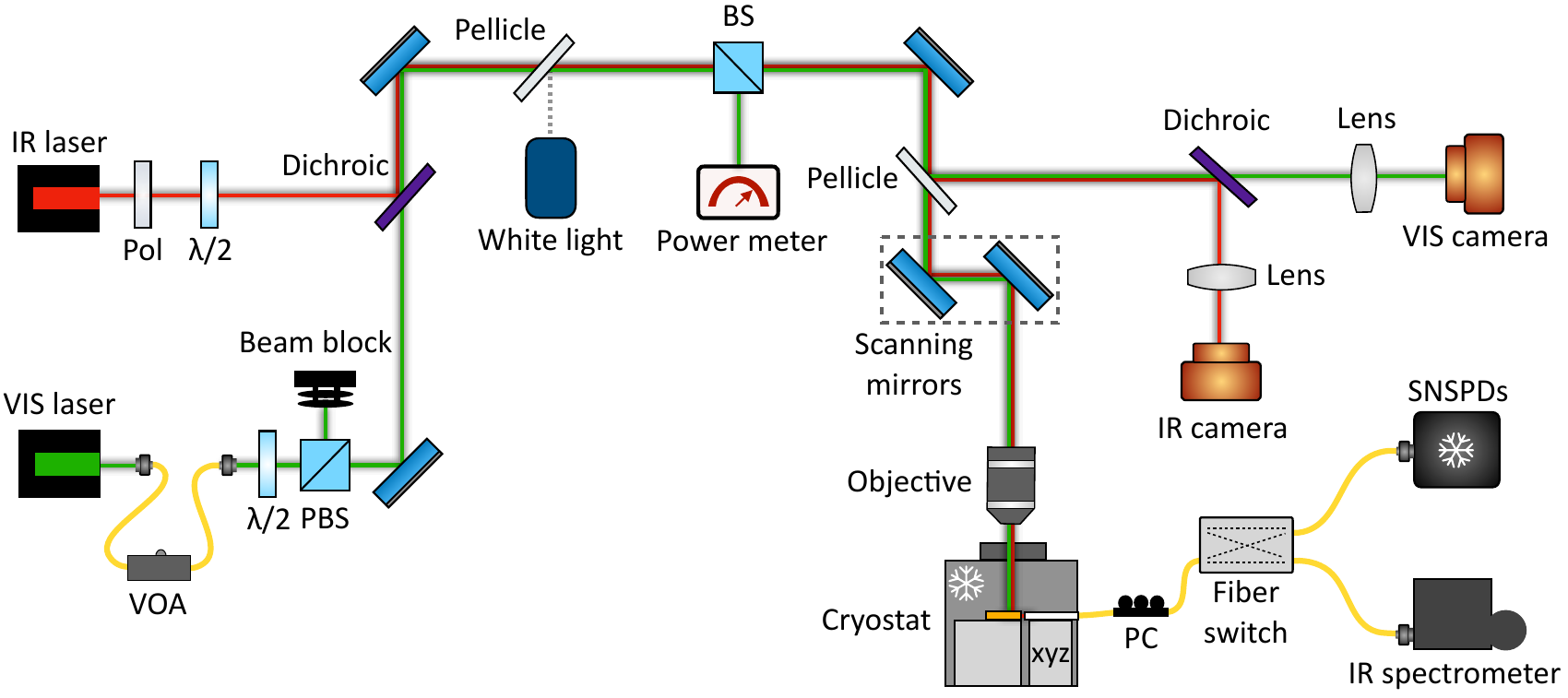}
    \caption{Measurement setup schematic. Our setup consists of a cryogenic confocal microscope equipped with a homemade fiber feedthrough. Infrared (IR) and visible (VIS) laser beams pass through polarization and power control components and are combined at a dichroic mirror to focus onto a sample placed into a cryostat. Their position on the sample is controlled by scanning mirrors, and their focus is set by an objective. A polarizer (Pol) and a half-wave plate ($\lambda/2$) are used on the IR path, while a variable optical attenuator (VOA), $\lambda/2$ and a polarizing beam splitter (PBS) are placed on the VIS path. A beam splitter (BS) is used to read the VIS laser power with a power meter. The sample is placed next to a movable fiber, which collects the PL. The PL is then routed either into superconducting nanowire single-photon detectors (SNSPDs) or an IR spectrometer via a fiber switch, after polarization adjustment via polarization controllers (PC). The IR and VIS reflections from the surface of the sample are imaged on an IR or VIS camera, respectively. White light is used to image the surface of the sample.}
    \label{fig:setup}
\end{figure*}

Our measurement setup is illustrated in Supplementary Fig.~\ref{fig:setup}. Two continuous-wave (CW) laser beams, infrared (IR) and visible (VIS), are combined at a dichroic mirror to shine onto the sample after passing through a set of scanning mirrors and an objective (Mitutoyo 50× M Plan APO). Polarization components such as a polarizer (Pol), a half-wave plate ($\lambda/2$), and a polarizing beam splitter (PBS) are used for polarization and power adjustments. Additionally, a variable optical attenuator (VOA) from Thorlabs is used for automated sweeps of the VIS laser power. A beam splitter (BS) and a power meter are placed in the setup to read out the VIS laser power. The VIS beam is a Coherent Verdi G5 at $532$~nm and is used for G-center excitation, while the IR laser is a tunable laser TSL-570 from Santec set to $1280$~nm and serves as a probe for sample alignment. The photoluminescence (PL) originating from the excitation of our waveguide-integrated G-centers is collected into a fiber aligned to the sample waveguide. The ultra-high numerical-aperture fiber is mounted on XYZ cryogenic piezoelectric stages from Attocube and placed next to the sample into the cryostat (Montana Instruments CR-057). After polarization adjustment via polarization controllers (PC), the PL is routed to either superconducting nanowire single-photon detectors (SNSPDs) from Photon Spot or to an IR spectrometer from Princeton Instrument, via the use of a fiber switch from Photonwares. Our SNSPDs feature detection efficiencies of up to 24 \%, and are readout with a Swabian Instruments Timetagger 20. Our IR spectrometer consists of a PyLon IR CCD array and two different gratings, one with a density of $300$~gr/mm and a $1.2$ \textmu m blaze and another with a density of 900 gr/mm and a $1.3$ \textmu m blaze. They lead to pixel-defined resolutions of $155$~pm and $40$~pm, respectively. The IR and VIS reflections from our sample are first separated with a dichroic mirror and then collected with a VIS camera from Thorlabs and an IR camera from Allied Vision, respectively, for laser imaging. White light is used to image the surface of the sample. Excited-state lifetime measurements of our G-centers are performed with a pulsed laser (SuperK from NKT Photonics) with a maximum repetition rate of 78 MHz and filtered by a bandpass filter centered at $532$~nm. For PL measurements, a free-space bandpass filtering setup (not shown in the figure), composed of a $1250$~nm longpass filter and a $1300$~nm shortpass filter, is used to isolate a $50$~nm-wide region, including the ZPL. 
To measure the second-order correlation function and excited-state lifetime, the PL is instead filtered with a tunable wavelength filter from WLPhotonics with a bandwidth of $0.1$~nm.

\section{Shadowmask metal evaporation}
\label{sec:sdm}

In this work, shadow mask evaporation is used to deposit electrical pads directly onto the chip, with chromium ($50$ nm) and gold ($200$ nm) serving as the contact materials. This technique avoids both liftoff, which could risk structural failure of the suspended waveguide, and dry metal etching, which may lead to contamination of the sample. By using a shadow mask, we minimize both the risk of structural collapse and potential chemical contamination.
The shadow mask was fabricated from a $0.1$ mm ($4$ mil) thick steel sheet using an LPKF ProtoLaser U4 laser cutter at MIT’s T.J. Rodgers Laboratory. The laser cutter produces precise patterning, achieving a smallest hole radius of $25$ µm, with a minimum spacing of $40$ µm between holes. The hole radius used to pattern the electrical pads was 50 µm. The alignment accuracy of the shadow mask on the chip was within $25$ µm across the entire chip area ($3.7$ mm $\times$ $5.2$ mm), granting precise positioning of the pads. While very effective for our specific application, it is important to acknowledge that the alignment requirements of this technique would present limitations for achieving tight tolerances or for very large-scale evaporations.

For the evaporation setup, the shadow mask was positioned approximately $100$ µm above the chip and held in place by a steel structure (shown in Supplementary Fig.~\ref{fig:sdm}a), which stabilized the mask during deposition. After alignment, the holder assembly was placed into an electron beam evaporator (as illustrated in Supplementary Fig.~\ref{fig:sdm}c), where the metal pads were deposited via the mask without rotation applied to the chip holder. This configuration prevents pad shapes from broadening, resulting in better edge definition and deposition accuracy. After evaporation, the chip was liberated from its holder. Metal pads were wire-bonded to a printed circuit board before being placed in the experimental setup for characterization.

\begin{figure*}[tbhp]
	\centering
	\includegraphics[width=\linewidth]{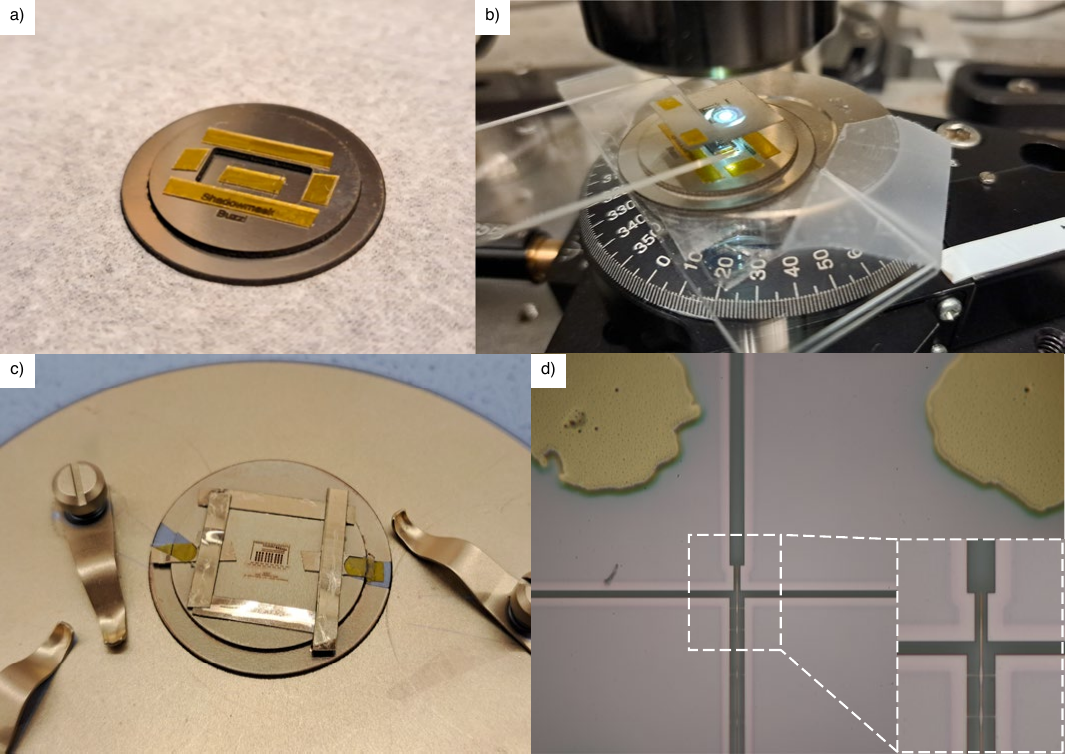}
    \caption{\textbf{Alignment of the steel shadow mask for electrical pad deposition.} 
    \textbf{a)} Steel structure used to position the mask approximately 100 µm above the chip. 
    \textbf{b)} Setup for aligning the steel shadow mask with the chip. The mask is aligned with an optical microscope to ensure that the mask holes are accurately positioned over the designated pad areas on the chip. 
    \textbf{c)} Entire assembly with the holder and aligned mask, shown here after the electron beam evaporation of the chromium-gold pads through the mask.
    \textbf{d)} Optical microscope image of a fabricated device, displaying the waveguide cantilever and the metal pads deposited on the chip surface. The inset shows a magnified view of the cantilever structure.}

    \label{fig:sdm}
\end{figure*}

\section{Tether loss and coupling efficiency}

\begin{figure*}[tbhp]
	\centering
	\includegraphics[width=\linewidth]{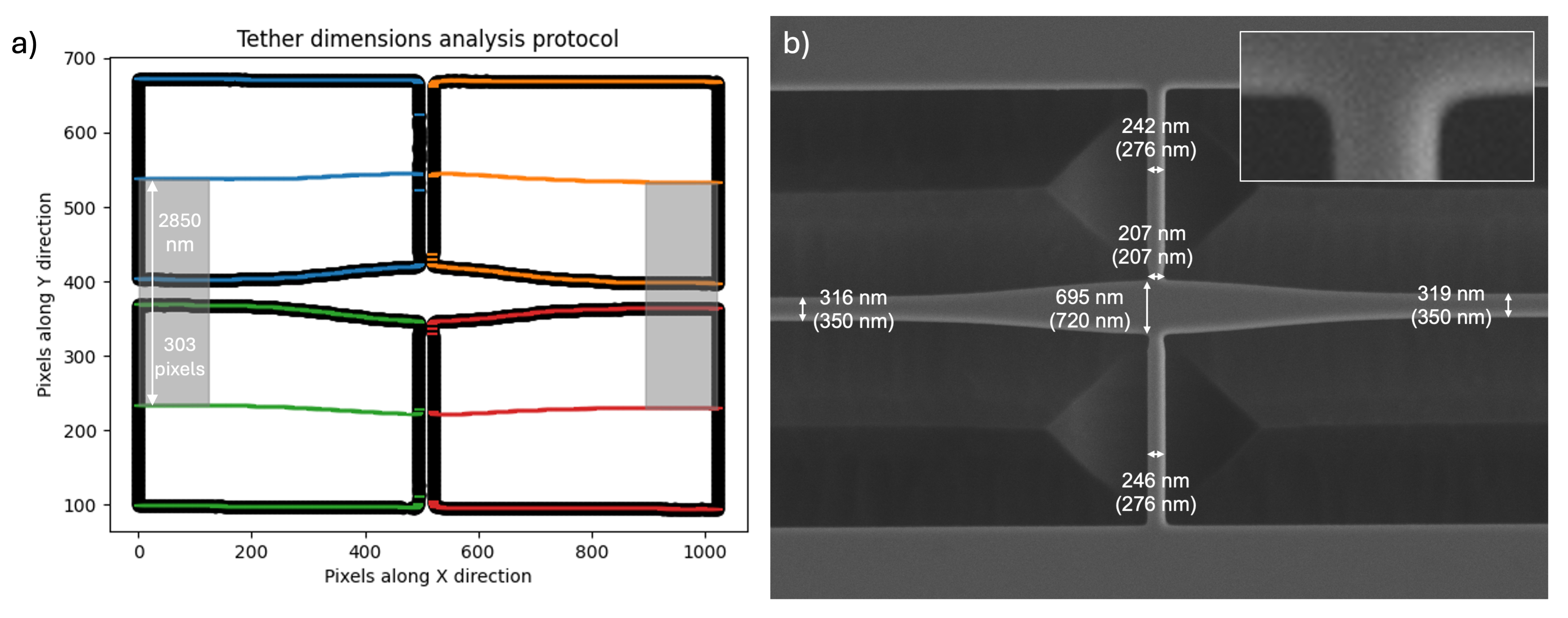}
    \caption{Tether dimensions. \textbf{a)} Protocol to retrieve the actual tether dimensions from an SEM image. The contours are extracted using the \texttt{cv2} Python module. Due to the working principle of the electron beam lithography and reactive ion etching, the distance between the center of the upper and lower cut-out region can be assumed to be equal to the designed $2850$~nm. If too much (or too little) material is removed, this is supposed to happen equally on all edges, and hence, the location of the center is not altered. Averaging over the gray area indicated in the figure, this $2850$~nm center separation corresponds to 303 pixels. \textbf{b)} SEM image of the tether with indications of the estimated (designed) dimensions. The inset shows the pixel resolution. The pixel size of $9.4$~nm is a good indication of the measurement uncertainty in this protocol.}
    \label{fig:tether}
\end{figure*}

The tether dimensions of the fabricated devices differ slightly from the designed values and are extracted from an SEM image (Supplementary Fig.~\ref{fig:tether}). Another FDTD simulation using the extracted tether dimensions predicts a transmission efficiency $\eta_{\text{tether}}$ of $94$~\% for the quasi-TE mode at $1280$~nm, shown in Supplementary Fig.~\ref{fig:TEtether}. Experimentally, the in/out-coupling efficiency is determined by sending $1280$~nm light through a loop-back suspended waveguide consisting of 15 tethers. The characterization is done using an array of UHNA3 fibers that are spliced to SMF28 fibers to connect to the experimental setup (cf. Supplementary Note~\ref{sec:splice}). The splice of the input (output) fiber has a transmission efficiency $\eta_{\text{splice,in}}$ ($\eta_{\text{splice,out}}$) of $88$~\% ($80$~\%). For an input power $P_\text{in}$ of 
 $1.0\ \text{mW}$ the output power $P_\text{out}$ equals $4.2\ \text{\textmu W}$. Assuming the tether loss is the main contribution to the propagation loss, the coupling efficiency $\eta_{\text{fiber-taper}}$ of the inverse linear taper to the UHNA3 fiber can be extracted from $P_\text{out} = P_\text{in} \eta^{15}_{\text{tether}} \eta^2_{\text{fiber-taper}} \eta_{\text{splice,in}} \eta_{\text{splice,out}}$. This results in $\eta_{\text{fiber-taper}}=12\%$, which combines the taper efficiency, mode matching between the fiber and the taper tip, and Fresnel reflections at the interface. The measured value represents a lower bound since the setup did not allow for optimizing the pitch, roll, and yaw angles.

\begin{figure*}[tbhp]
	\centering
	\includegraphics[]{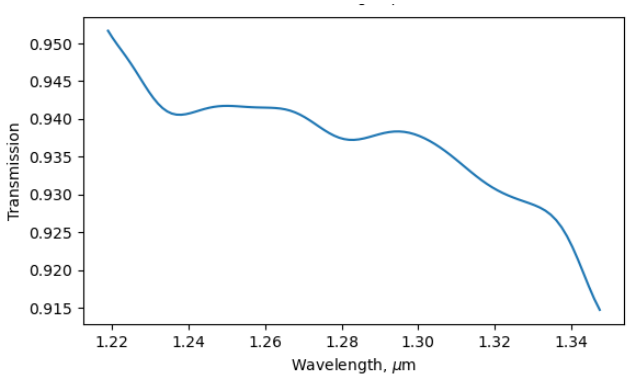}
    \caption{FDTD simulation of the transmission of the fundamental quasi-TE waveguide mode. The dimensions are set according to the fabricated and measured values.}
    \label{fig:TEtether}
\end{figure*}

\section{Additional spectroscopy results }
\subsection*{Above-bandgap excitation of G-centers}

\begin{figure*}[tbhp]
	\centering
	\includegraphics[width=\linewidth]{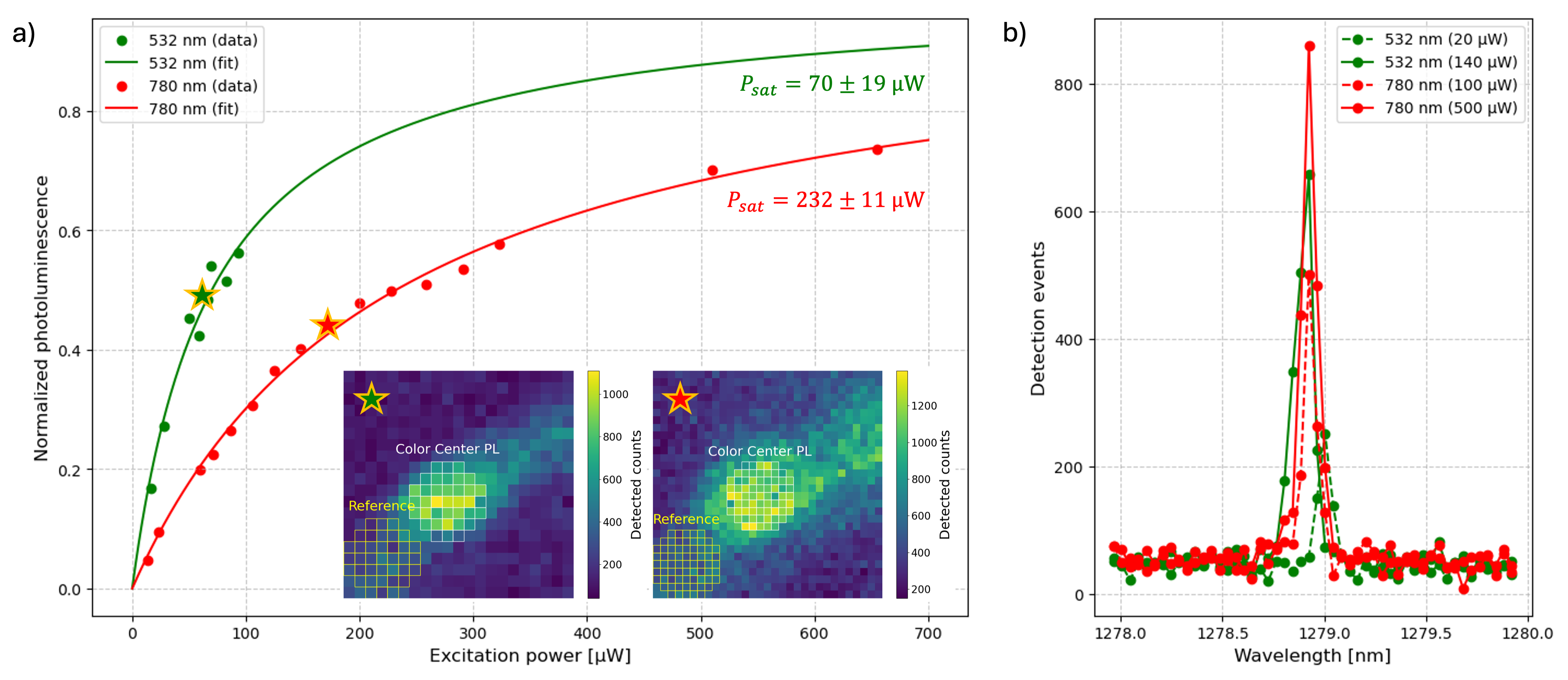}
    \caption{Above-bandgap excitation of G-centers with a $532$~nm and $780$~nm laser. \textbf{a)} Power saturation curves. 2D photoluminescence raster maps are made using an SNSPD while sweeping the confocal excitation power (cf. inset of with corresponding stars). The effective photoluminescence count rate is extracted from integrating the counts on the scans around the color center location and subtracting the background counts taken at a reference location in the same waveguide. The background-corrected photoluminescence is fitted to a 2-level emitter saturation model: $I(P)=I_{\text{max}}\frac{P}{P+P_{\text{sat}}}$ and normalized to the fitted $I_{\text{max}}$. \textbf{b)} Spectra of G-center photoluminescence below and above saturation power for both excitation wavelengths.}
    \label{fig:confocal}
\end{figure*}

To demonstrate our ability to excite G-centers in suspended structures, their photoluminescence is characterized in a simple loop-back waveguide, using a $532$~nm and $780$~nm laser for above-bandgap excitation. The chip containing this device is fabricated in the same way as the sample described in the main text of this paper but with a 10 times lower implantation dose ($5\times 10^{12}$ ions/cm$^2$) of a different carbon isotope ($^{13}$C) and without electrical pads. The light emitted by the color center is collected using a UHNA3 fiber that is edge-coupled to one of the tapered ends of the suspended waveguide. 2D photoluminescence raster maps are acquired as described in the experimental methods of the main text. The background-corrected emission intensity is extracted for various excitation powers and fitted to a 2-level power saturation model (Supplementary Fig.~\ref{fig:confocal}a). The extracted saturation power is $70 \pm 19$ \textmu W for excitation at $532$~nm and $232 \pm 11$ \textmu W at $780$~nm. This discrepancy might arise from different focal spots on the suspended waveguide for both beams and also from the contribution of a higher absorption coefficient at $532$~nm compared to $780$~nm (approximately 7 times larger). Photoluminescence spectra are characterized below and above saturation power (Supplementary Fig.~\ref{fig:confocal}b). The intensity of the emitted light is proportional to the integrated counts of the peak in the spectrum after subtracting the background counts. Taking into account that $I(140$ \textmu W$)=0.67I_{\text{max}}^{532\text{nm}}$ and $I(500$ \textmu W$)=0.68I_{\text{max}}^{780\text{nm}}$ for excitation at the respective wavelengths and powers, we find that $\frac{I_{\text{max}}^{780\text{nm}}}{I_{\text{max}}^{532\text{nm}}}=1.13\pm 0.10$, after propagating errors from the saturation power fit. Since the difference between the calculated ratio and unity is less than $2\sigma$, we conclude that the maximum intensities achieved under the two excitation conditions are not significantly different.

\subsection*{Additional lifetime, second-order correlation and saturation curves}
In the following section, we provide additional results from spectroscopy measurements performed on G-centers on localization $A$ and $B$, all performed at $\lambda_{exc}=532$ nm. In Supplementary Fig.~\ref{fig:spectroB}(a), we provide the time-resolved photoluminescence upon pulsed above-band excitation at low power. The data is fitted to a mono-exponential decay, from which we extract a total decay timescale of  $\tau = (6.4\pm0.1)$~ns. This value is similar to the value reported in the main text for $A_2$ and to previous work \cite{saggio_cavity-enhanced_2024,prabhu_individually_2023} and has been shown to be excitation power-independent \cite{prabhu_individually_2023}. We further provide the saturation curve of the filtered ZPL $A_2$ under pulsed above-band excitation, displayed in Supplementary Fig.~\ref{fig:spectroB}(b). A fit to the saturation of a two-level system provides a saturation power of $P_\text{sat}=(3\pm2)$ µW and a saturated filtered intensity of $I_{\inf}=(178\pm61)$ counts/s. The lifetime presented in the main text is measured close to saturation. We continue the characterization by measuring the saturation of the same line in continuous-wave excitation, as shown in Supplementary Fig.~\ref{fig:spectroB}(c), from which we extract a saturation power of $P_\text{sat}=(11.5\pm2.7)$ µW and saturated filtered intensity of $I_{\inf}=(405\pm32)$ counts/s. We then park the laser power around saturation and measure the second-order correlation function,  $g^{(2)}(\tau)$, shown in Supplementary Fig.~\ref{fig:spectroB}(d). Due to the low coincidence count rate, the fit to a three-level system, which includes a bunching term, results in high fitting errors. We then show the fit to the second-order correlation function of a two-level system. The anti-bunching time constant is extracted as $\tau_a= 0.95 \pm 0.65 $ ns, however it is not grasping the full physics of the system since slight bunching at short time delays is visible. Nonetheless, we can report an experimentally measured value of $g^{(2)}(0)=0.12\pm0.05$, validating the hypothesis of a single quantum emitter. The error on $g^{(2)}(0)$ is estimated from Poissonian statistics.
 \subsection*{Broadband G-center spectra}
 A specific attribute to genuine G-center is the presence of a red-detuned emission line due to local vibrational mode of the defect \cite{thonke_new_1981}, around 1381 nm, together with a zero-phonon line around 1278 nm. We record such a broadband spectrum by removing all filtering, shown in Supplementary Fig.~\ref{fig:fig_rev_Eline}(a). The spectra are recorded from a waveguide-coupled G-center originating from the same die as strain tuning results obtained in Supplementary Fig.~\ref{fig:df2}. Both emission lines, the zero-phonon line, G-line (Supplementary Fig.~\ref{fig:fig_rev_Eline}(b)), and the E-line (Supplementary Fig.~\ref{fig:fig_rev_Eline}(c)), originate from the same excited state, we therefore expect similar saturation behavior. We record the spectrum as a function of excitation power and extract the integrated intensity at each peak. We then fit the integrated intensity as a function of power, which yields a similar saturation power for both emission lines, $P_\text{sat}\approx122 \,\mu$W, as shown in Supplementary Fig.~\ref{fig:fig_rev_Eline}(d).  We note that the energy detuning between the E-line and the ZPL measured here (73.6 meV) differs slightly from reported values in the literature \cite{thonke_new_1981,aberl_all-epitaxial_2024} (71.9 meV and 71.6 meV). In our experiment, we report a large intensity ratio between the E-line and ZPL, which resembles experimental values reported in Ref.~\cite{thonke_new_1981} and is compatible with results from ab-initio calculations \cite{aberl_all-epitaxial_2024}. This high intensity ratio may be facilitated by the broadband coupling efficiency, differing from Ref. \cite{durand_genuine_2024}.
 \begin{figure*}[h!]
    \centering
    \includegraphics[scale=0.9]{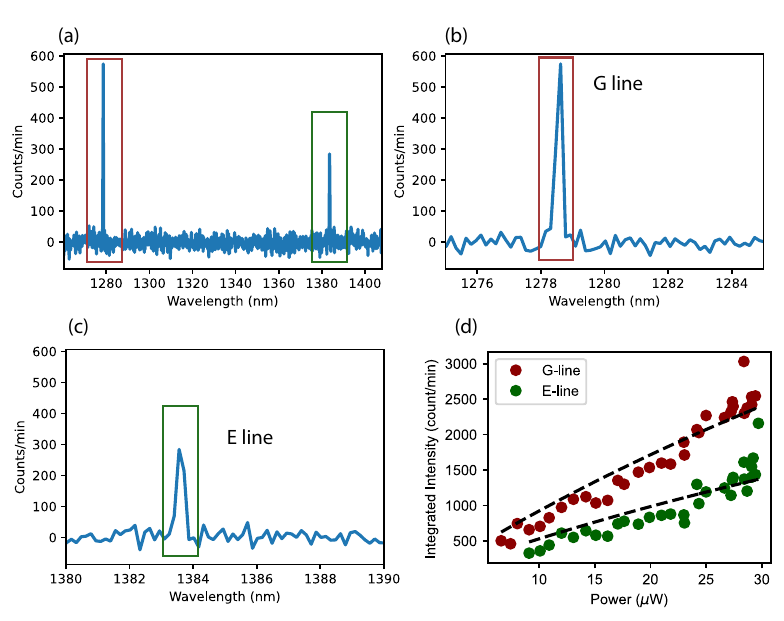}
    \caption{(a) Broadband spectrum of waveguide-coupled G-center emission under above-band excitation. The spectrum reveals  (b) the zero-phonon line and (c) the E-line, originating from local vibrational modes and expected around 1381 nm. (d) Through power-dependent measurement, we verify that both emission lines originate from the same excited state, by showing similar saturation behavior with $P_\text{sat}\approx122\ $\textmu W. }
    \label{fig:fig_rev_Eline}
\end{figure*}

\begin{figure*}[tbhp]
	\centering
	\includegraphics[]{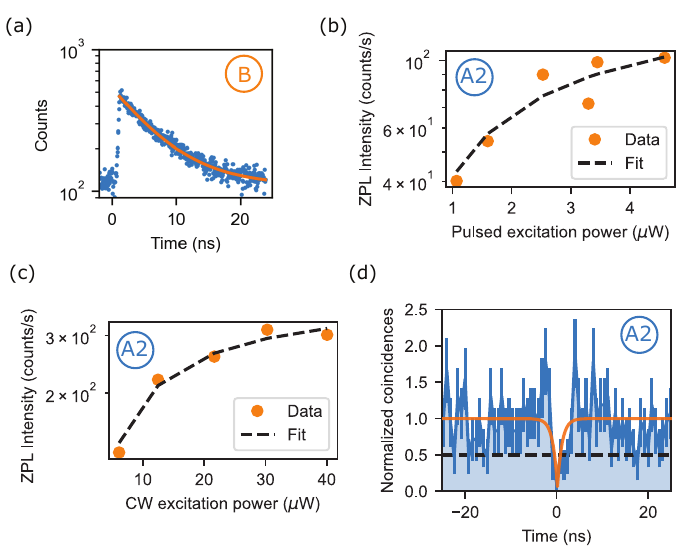}
    \caption{\textbf{a)} Time-resolved measurement of the PL from the color center localized on position $B$. The fit to a mono-exponential decay leads to $\tau = (6.4\pm0.1)$~ns. \textbf{b)} Filtered ZPL from $A_2$ as a function of pulsed above-band excitation power. The dashed line indicates the fit to power saturation of a two-level system, from which we extract a saturation power of $P_\text{sat}=(3\pm2)$ µW. \textbf{c)} Filtered ZPL from $A_2$ as a function of CW above-band excitation power. From the fit (dashed line), we extract $P_\text{sat}=(11.5\pm2.7)$ µW. \textbf{d)} Second-order correlation function of $A_2$, excited in continuous-wave around saturation, with a measured $g^{(2)}(0)=0.12\pm0.05$. The orange line is a fit for a two-level system. The dashed black line indicates the threshold for interaction with a single emitter.}
    \label{fig:spectroB}
\end{figure*}
\subsection*{Spectroscopy as a function of tuning voltage}
\begin{figure*}[tbhp]
    \centering
    \includegraphics[width=\linewidth]{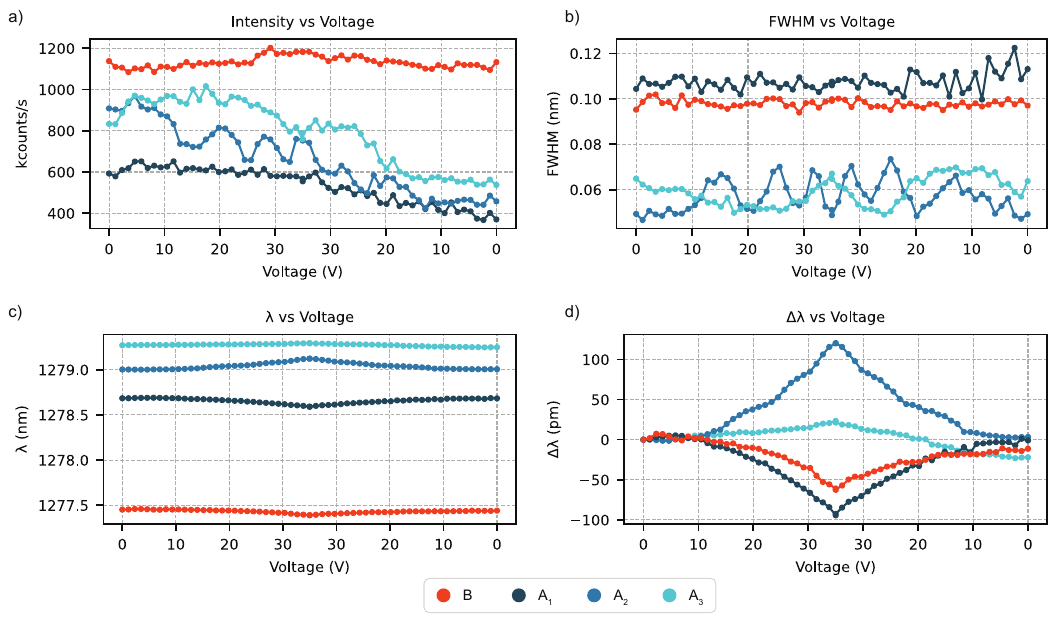}
    \caption{Characteristics of the color centers during tuning (voltage up and down). \textbf{a)} Intensity extracted from a fit to a Lorentzian as a function of applied voltage. \textbf{b)} Full width at half maximum (FWHM) of the Lorentzian fit as a function of voltage. \textbf{c)} Color center central emission wavelength versus voltage. \textbf{d)} Shift in color center central emission wavelength versus voltage.}
    \label{fig:tun}
\end{figure*}

Figure~\ref{fig:tun} presents the full characteristics of the color centers during the tuning process. Supplementary Fig.~\ref{fig:tun}a shows the intensity, computed as the amplitude of the Lorentzian fit in this case, as a function of voltage. The observed decrease in intensity for lines $A_1$, $A_2$, and $A_3$ is attributed to the gradual misalignment of the excitation laser during measurements caused by the relaxation of the galvos. However, the laser spot was optimized at the start of each measurement series to reduce the misalignment issue. Such a drift was not clearly observed for the $B$ line because its spot position was intentionally aligned with the galvos' relaxation position. Supplementary Fig.~\ref{fig:tun}b illustrates the full width at half maximum (FWHM) of the emission peaks versus voltage. Supplementary Fig.~\ref{fig:tun}c depicts the central emission wavelength of the color centers as a function of voltage. Lastly, Supplementary Fig.~\ref{fig:tun}d shows the shift in the central emission wavelength versus voltage. The oscillations visible in Supplementary Fig.~\ref{fig:tun}d, particularly visible for emitter \(A_2\), result from a small dark region between the pixels of the spectrometer. This phenomenon affects the Lorentzian fitting. From Fig.~3b of the main text, we observe that these oscillations align with the pixel centers of the spectrometer, with dips corresponding to when the emission wavelength is tuned precisely between two pixels. We note that the strain tuning mechanism is reversible, reproducible, and does not affect the linewidth of the emitter. Finally, we provide in  Supplementary Fig.~\ref{fig:iv} the current-voltage map recording during the actuation, which reveals a maximum power dissipated on the order of $10$~nW, demonstrating the low-power operation of the device.

\begin{figure*}[tbhp]
    \centering
    \includegraphics[]{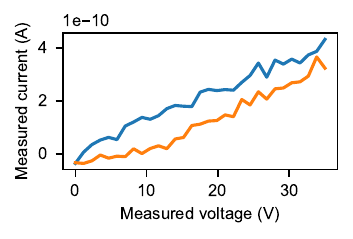}
    \caption{Measured current as a function of the applied voltage between the cantilever and the substrate, recorded during the tuning of the color centers. The blue curve is the forward sweep, while the orange curve is the backward sweep. The maximum power dissipated in this operation is $\approx 10$~nW.}
    \label{fig:iv}
\end{figure*} 
\section{UHNA3-SMF28 fiber splicing optimization} \label{sec:splice}
To maximize the coupling efficiency between the taper edge couplers and a single mode fiber, we employ an array of UHNA3 fibers with a mode field diameter of $3.3 \pm 0.3$ µm at $1310$ nm (according to Thorlabs website). Without any additional care, however, the coupling to SMF28 fibers, which constitute the rest of our fiber network, would suffer from loss due to the differing numerical aperture (NA), such that $a=\text{NA}_\text{UHNA}/\text{NA}_\text{SMF28}=0.35/0.14$ leads to a maximal coupling efficiency of $\eta=4a^2/(1+a^2)^2\approx48\%$. To maximize the coupling efficiency and hence maximize the single-photon count rate, we splice the UHNA3 end fiber from the array with an SMF28 fiber following the recipe provided in Refs. \cite{yin_low_2019,du_demonstration_2024}. We optimize the arc time during the splice by measuring the power of a continuous-wave laser at $1280$~nm at the output of the fiber array with a power meter head and normalize it by the input power. As shown in Supplementary Fig.~\ref{fig:splice}, the initial splicing efficiency is improved by approximately $50\ \%$ after further arc splice.
\begin{figure*}[tbhp]
	\centering
	\includegraphics[]{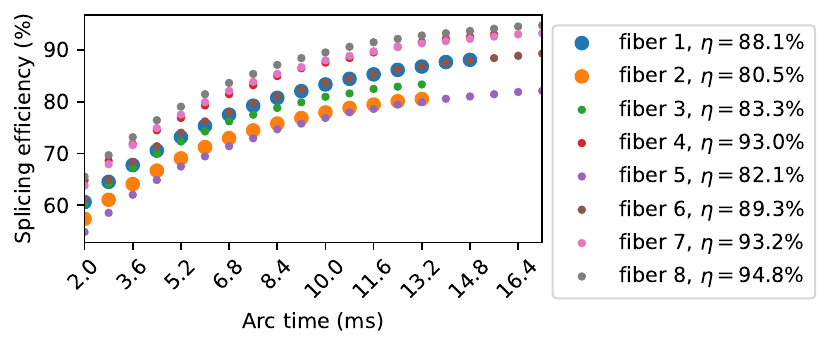}
    \caption{Splicing optimization between the UHNA3 array fibers and SMF28. The transmission efficiency is monitored for each extra arc. The fibers used in this work are highlighted with larger markers (fiber $1$, fiber $2$). The final transmission efficiencies after splicing are indicated in the legend.}
    \label{fig:splice}
\end{figure*}

\section{Failure mode of the cantilever actuation}

During the measurement, we observed the failure of the device due to the collapse of a detached part of the lateral pad onto the cantilever, as shown in Supplementary Fig.~\ref{fig:df1}a, where the entire device is visible. The collapse itself is more clearly seen in the inset in Supplementary Fig.~\ref{fig:df1}b, where the portion of the lateral pad that collapsed onto the cantilever waveguide is highlighted, causing structural damage and leading to the device’s failure. Supplementary Fig.~\ref{fig:df1}c shows a simulation of the cantilever’s maximum displacement as a function of applied voltage, illustrating the transition from 0 V to the pull-in voltage ($42.5$~V).

In a different device, with a cantilever length of $15$~\textmu m, shown in Supplementary Fig.~\ref{fig:df2}, the heatmap (Supplementary Fig.~\ref{fig:df2}a) illustrates the tuning of the emitters as a function of the vertical driving voltage. At $12.5$~V, the right cantilever collapsed onto the central part of the lateral pad (Supplementary Fig.~\ref{fig:df2}b). Initially, the emitters tuned as anticipated, but after the collapse, the device exhibited approximately linear tuning at higher voltages due to the continued possible driving of the collapsed cantilever. The failure occurred because the pad was left floating, and likely, the charging exceeded the lateral pull-in voltage, causing the collapse. Interestingly, a large tuning exceeding 300 pm was observed at $12.5$~V. Moreover, we can see from the heatmap in Supplementary Fig.~\ref{fig:df2}a that two color centers on the right were tuned at the same frequency before the collapse, at approximately $10$~V. The presence of two distinct emission lines after the failure confirms that the tuning observed prior to collapse originated from two separate emitters.

\begin{figure*}[tbhp]
	\centering
	\includegraphics[width=\linewidth]{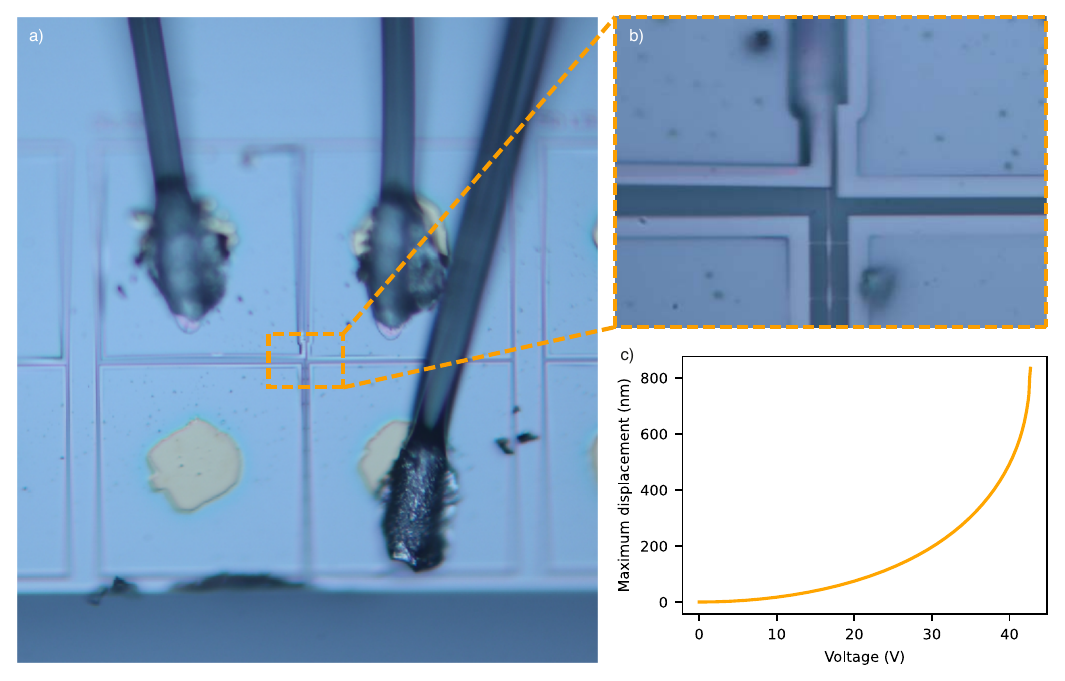}
    \caption{Post-failure device structure. \textbf{a)} Image of the entire device, showing the deposited electrical pads and wire bonds on the chip. The device is observed after the mechanical failure.  
    \textbf{b)} Optical microscope close-up of the cantilever waveguide. The image highlights the cause of the failure, revealing a portion of a cracked lateral pad that has collapsed onto the cantilever waveguide, leading to structural damage.  
    \textbf{c)} FEM-simulated displacement of the cantilever waveguide as a function of applied voltage, showing the curve from $0$~V to the pull-in voltage ($42.5$~V).}
    \label{fig:df1}
\end{figure*}

\begin{figure*}[tbhp]
	\centering
	\includegraphics[width=\linewidth]{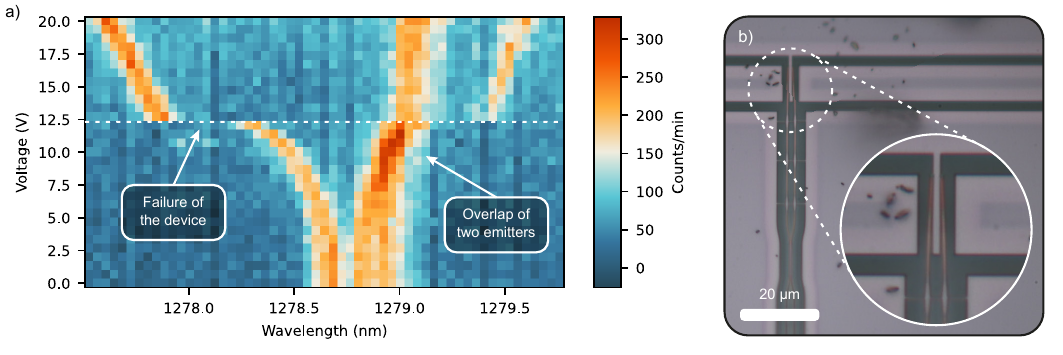}
    \caption{Tuning of emitters up to device failure. \textbf{(a)} Heatmap showing the emission spectrum as a function of driving voltage. At $12.5$~V, the device collapses laterally. \textbf{(b)} Optical microscope image of the device after failure, illustrating the cantilever lateral collapse.}
    \label{fig:df2}
\end{figure*}

\section{Secondary Ion Mass Spectroscopy}

Secondary ion mass spectroscopy (SIMS) was performed by Eurofins EAG Materials Science, LLC after annealing and chip fabrication on an edge piece from the same sample as the devices. The results for carbon, silicon, and oxygen concentrations as a function of depth are shown in Supplementary Fig.~\ref{fig:sis}. The increase in oxygen concentration reveals the interface between the silicon device layer and the bottom silicon dioxide. The nanoscale localization model uses the SIMS data, considering only the carbon concentration in the central region of the waveguide thickness. Carbon concentrations at the surface and interfaces are excluded, as they are likely due to contamination during processing and do not contribute to the formation of G-centers. As described in the Methods section, the distribution of G-centers is assumed to be proportional to the square of the carbon concentration in the central region of the waveguide.

\begin{figure*}[tbhp]
	\centering
	\includegraphics[width=\linewidth]{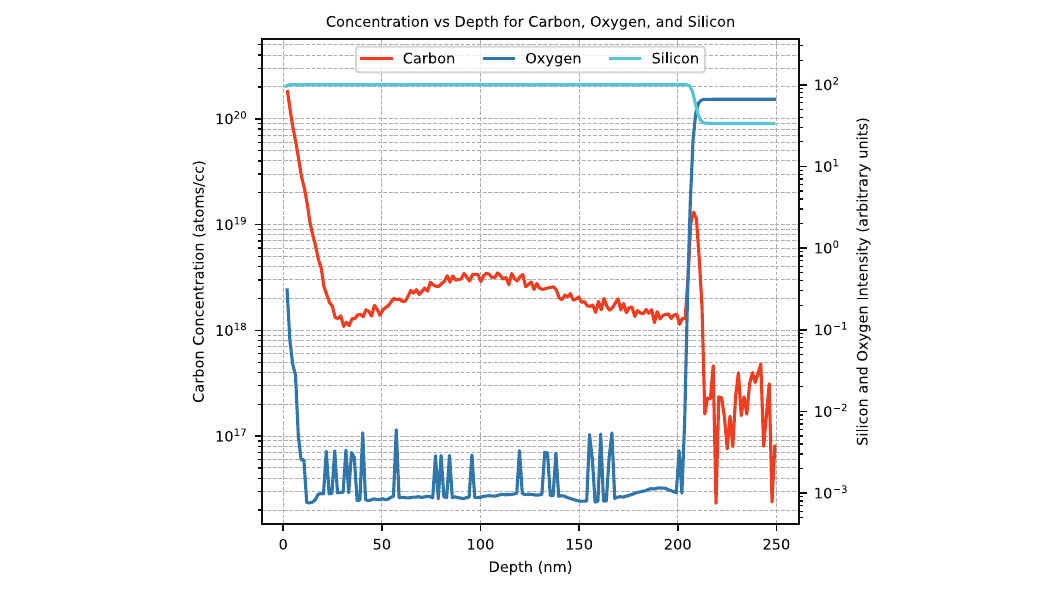}
    \caption{Secondary ion mass spectroscopy (SIMS) data of the sample, displaying the carbon concentration alongside silicon and oxide concentrations as a function of depth.}
    \label{fig:sis}
\end{figure*}

\section{Coupling efficiency to the waveguide mode}
The G-centers can be found with three different classes of dipole orientations. The dipole coupling to the waveguide mode described in Fig.4(f) in the main text is quantified by the $\beta$-factor, which is the ratio between the power coupled to a given waveguide mode and the total power radiated by the
dipole. Following the method in Ref. \cite{rotenberg_small_2017}, we calculate the power radiated in the fundamental quasi-TE waveguide mode
by performing a 3D FDTD \cite{taflove_computational_nodate} calculation carried out using MEEP open-source
software package \cite{oskooi_meep_2010}. To do so, we define a single-mode waveguide, which is $350$ nm wide and $220$ nm tall. The position of a radiating
dipole, either oriented along the $[110], [\overline{1}10]$ or $[011]$, is swept across the waveguide width and height, as shown in Fig.~4c-f in the main text.
The power coupled into the quasi-TE0 mode, whose mode profile is shown in Supplementary Fig.~\ref{fig:fdtd}(a) from Supplementary Note 1, is simulated and gives the $\beta$-factor for each equivalence class of dipoles.  We assume unidirectional emission thanks to the presence of the reflector, optimized for reflection of the quasi-TE0 mode (Supplementary Note 1, Supplementary Fig.~\ref{fig:reflector}). In the main text, we refer to this simulated quantity as the dipole coupling efficiency to illustrate the probability of detecting each dipole class given its position. This is a more accurate description of the value, since the $\beta$-factor is usually used to quantify the portion of quantum emitter radiation participating in coherent effects. However, in the case of G-centers and other color centers in silicon, the Debye-Waller, quantifying the amount of radiation emitted in the zero-phonon line instead of the phonon side band, is only of $F_{DW}=0.15$ \cite{beaufils_optical_2018} and will therefore represent an upper bound to the $\beta$-factor in the waveguide.
Additionally, we compute the coupling into the fundamental quasi-TM mode of the waveguide for each dipole class, as shown in Supplementary Fig.~\ref{fig:beta_TM}. We note that the $[110]$ dipole, due to its transverse orientation to the TM electromagnetic field distribution, couples poorly to this waveguide mode. Moreover, the $[\overline{1}10]$ dipole also shows low coupling to the TM waveguide mode, especially for small vertical offset. The value reported in Supplementary Fig.~\ref{fig:beta_TM} consider unidirectional emission, however the total coupling efficiency is weighted by the low reflection efficiency of the reflectors for the TM mode, as shown in Supplementary Fig.~\ref{fig:reflector}. To get the effective coupling efficiency, we need to apply a factor of $65$~\% to simulated values, leading all efficiencies to be less than 20 \%, further less if we restrict to dipoles located in central areas of the waveguide. Since the coupling efficiency to the TE waveguide mode is larger than to TM mode for all dipole classes, in the waveguide region of higher G-centers probability, we only consider the coupling to TE mode in the full model presented in Fig.4 in the main text.  

\begin{figure*}[tbhp]
	\centering
	\includegraphics[]{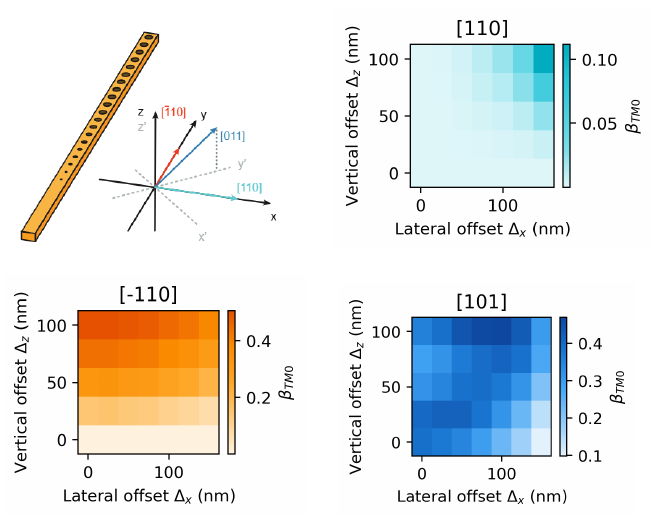}
    \caption{FDTD simulation of coupling efficiency to the quasi-TM waveguide mode for the three different classes of dipoles, as indicated in the schematics.}
    \label{fig:beta_TM}
\end{figure*}

\section{Nanoscale localization model for color centers}

The nanoscale localization model estimates the vertical positions and orientations of color centers within the cantilever waveguide. This supplementary note provides further details on the FEM simulations, the maximum likelihood estimation (MLE) of generation rates, and the Monte Carlo simulations. Additionally, we present the probability distributions for different scenarios, comparing results obtained with and without incorporating MLE for the color centers’ orientations.

\subsection*{FEM simulation and strain distribution}

The strain distribution along the cantilever waveguide was simulated using the electromechanical finite element method in COMSOL Multiphysics. The results are shown in Supplementary Fig.~\ref{fig:sim}a, where the heatmap illustrates how the maximum strain varies with different driving voltages along the cantilever. Supplementary Fig.~\ref{fig:sim}b presents the strain profile at a driving voltage of $35$~V, highlighting the spatial positions of the color centers.
\begin{figure*}[tbhp]
	\centering
	\includegraphics[width=\linewidth]{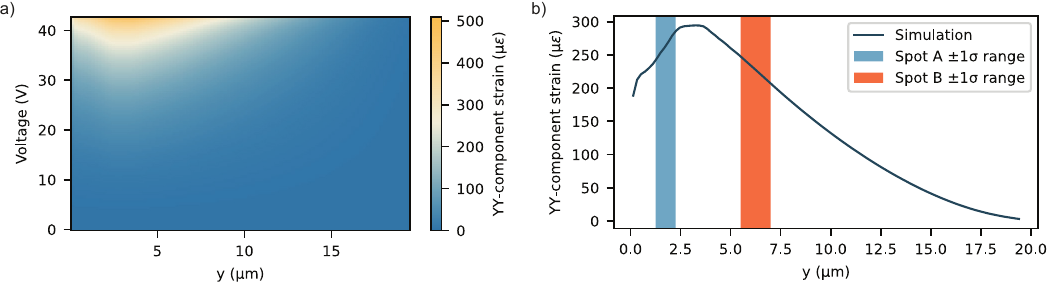}
    \caption{Strain distribution simulated by FEM along the cantilever. \textbf{a)} Heatmap of the strain distribution along the cantilever at different driving voltages. \textbf{b)} Strain profile along the cantilever at $35$~V, highlighting the spot positions corresponding to the color centers' positions along the cantilever.}
    \label{fig:sim}
\end{figure*} 
\subsection*{Maximum likelihood estimation of generation rates}

\begin{figure*}[tbhp]
    \centering
    \includegraphics[width=\linewidth]{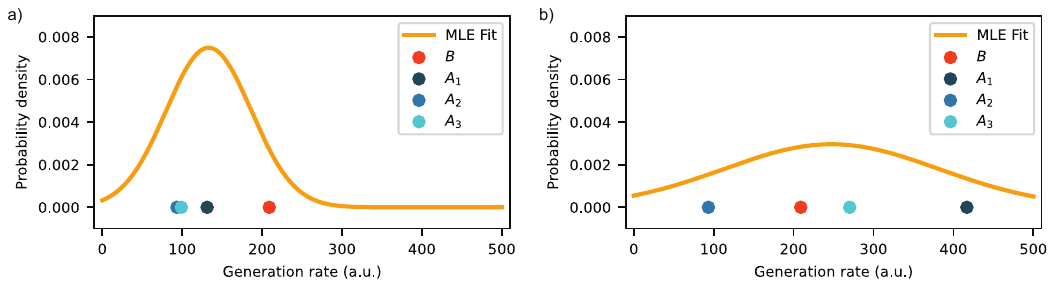}
    \caption{Maximum likelihood estimation of the generation rate for two scenarios with different dipole orientations of color centers. \textbf{a)} Estimation for the scenario where all emitters are aligned to the 0 equivalence class of orientations, corresponding to a more likely configuration. \textbf{b)} Estimation for the scenario where emitters $A_1$ and $A_3$ are aligned to the 1 equivalence class of orientations, representing a less likely configuration.}
    \label{fig:gen}
\end{figure*} 

As described in the main text, the generation rates for each emitter are estimated using MLE applied to Gaussian distributions. The MLE is performed on the estimated intensities of the emitters after accounting for coupling. The resulting values are proportional to the generation rates of the emitters. The intensities are computed by integrating the area under the Lorentzian fit of each peak and averaging over ten successive measurements. Supplementary Fig.~\ref{fig:gen} illustrates examples of likely and unlikely generation rate distributions based on this analysis.

\subsection*{Probability distributions for vertical positions and orientations}

\begin{figure*}[tbhp]
	\centering
	\includegraphics[width=\linewidth]{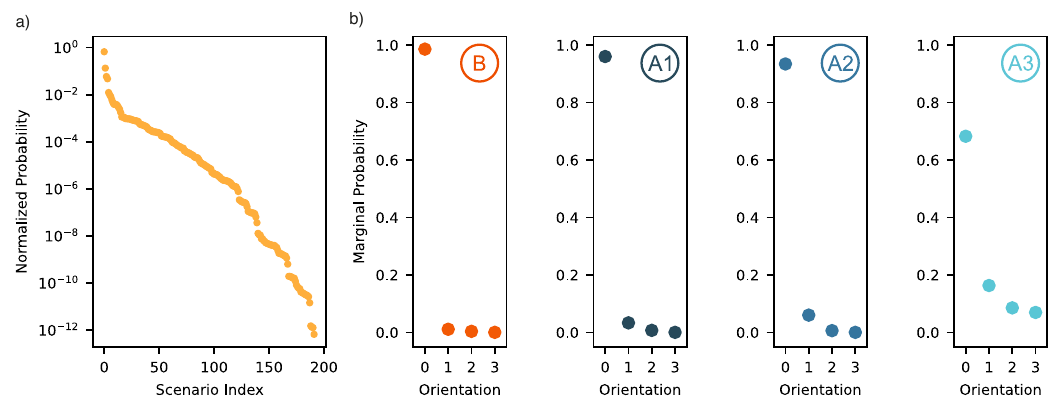}
    \caption{Probabilities of different dipolar orientation scenarios for each emitter. \textbf{a)} Joint probability distribution for the scenarios. \textbf{b)} Marginal probabilities for each emitter.}
    \label{fig:sce}
\end{figure*} 

To evaluate the probabilities of emitters occupying specific vertical positions and orientations, we calculate the product of two probabilities:
\begin{enumerate}
    \item The probability of an emitter being at a certain vertical position given its orientation, based on SIMS carbon concentration data.
    \item The MLE-derived probabilities of color centers aligning along specific dipole orientations.
\end{enumerate}

Supplementary Fig.~\ref{fig:sce} displays both the joint and marginal probability distributions for the emitters’ orientations and corresponding vertical positions. The joint distributions (Supplementary Fig.~\ref{fig:sce}a) represent multiple orientation scenarios occurring simultaneously, while the marginal distributions (Supplementary Fig.~\ref{fig:sce}b) depict probabilities for individual orientations. The orientation classes considered in the model are listed as classes 0 to 3, corresponding to the classifications in Supplementary Note 13: $F_1$, $F_2$, $F_3$–$F_6$, and $F_4$–$F_5$.

\subsection*{Monte Carlo simulations}

\begin{figure*}[tbhp]
	\centering\includegraphics[width=\linewidth]{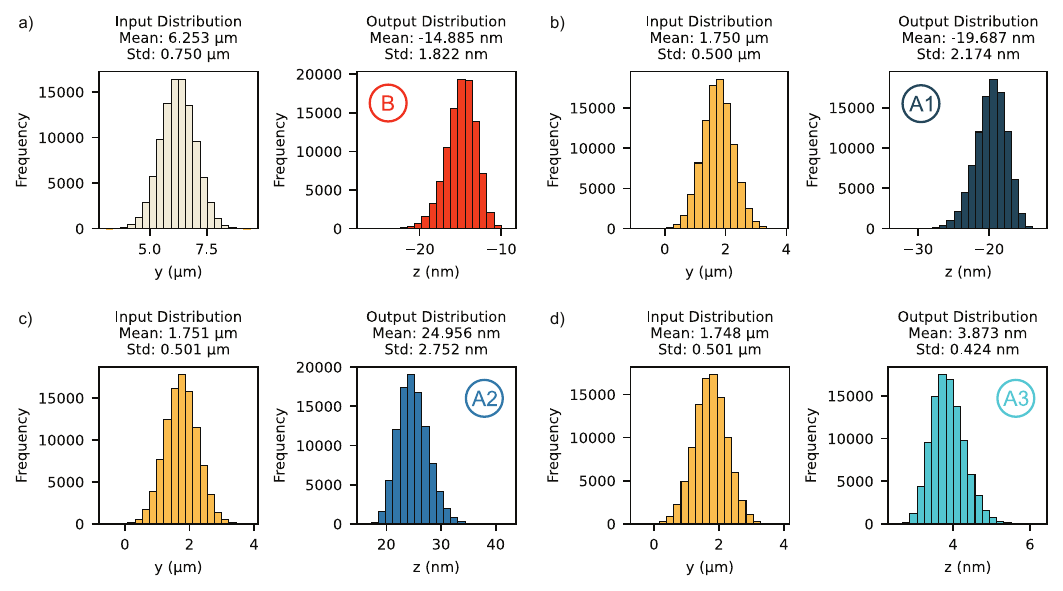}
    \caption{Results of Monte Carlo simulations estimating the error in the vertical localization of emitters. For each emitter, the left panels display the input probability distribution for the emitter's position along the cantilever, determined by the excitation spot position and size. The right panels show the output probability distribution, incorporating Gaussian errors of $20$~\% error from FEM simulations and $4.7$~\% error from piezospectroscopic constants. The mean and standard deviations of the distributions are reported on top of the plots. \textbf{a)} Emitter \( B \). \textbf{b)} Emitter \( A_1 \). \textbf{c)} Emitter \( A_2 \). \textbf{d)} Emitter \( A_3 \).}
    \label{fig:mon}
\end{figure*} 

Monte Carlo simulations were performed to quantify the uncertainty in vertical localization under the most likely orientation scenarios. The input distributions for the emitters’ vertical positions are shown in the left panels of Supplementary Fig.~\ref{fig:mon}, while the output distributions, incorporating Gaussian errors from FEM simulations ($20$~\%) and piezospectroscopic constants ($4.7$~\%), are displayed in the right panels. The mean and standard deviations of these distributions are reported at the top of each plot. Panels a) to d) refer to individual emitters $B$, $A_1$, $A_2$, and $A_3$, respectively.

\subsection*{Comparison of scenarios}

\begin{figure*}[tbhp]
	\centering
	\includegraphics[width=\linewidth]{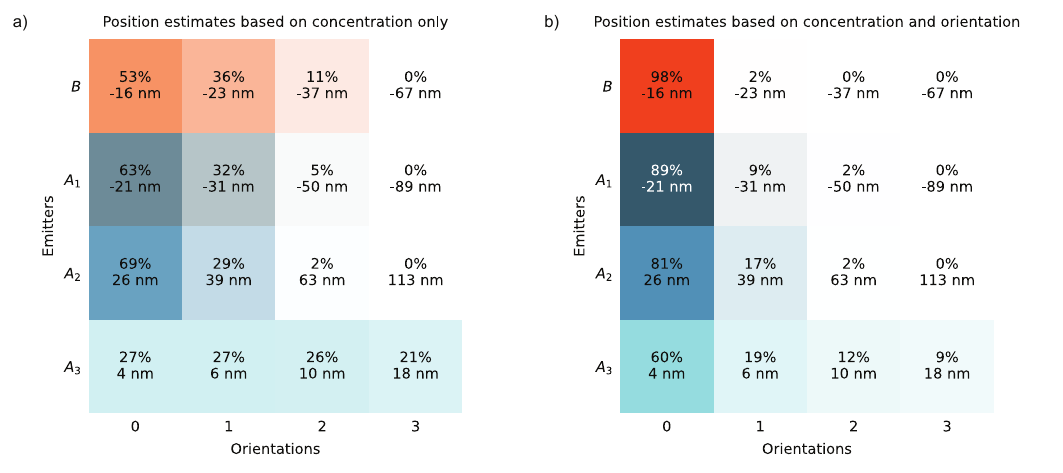}
    \caption{Distribution of probabilities for each vertical position of the color centers. \textbf{a)} On the left are the probabilities associated with each position, considering only the information based on the carbon concentration. \textbf{b)} On the right, the map shows the probabilities associated with considering both the carbon concentration and the probabilities given by the maximum likelihood estimation of the color centers' orientations.}
    \label{fig:pro}
\end{figure*} 

Supplementary Fig.~\ref{fig:pro} compares the probabilities of different scenarios. The left side of Supplementary Fig.~\ref{fig:pro}a presents probabilities derived from SIMS carbon concentration data only, assuming equal likelihood for all emitter orientations. The right side of Supplementary Fig.~\ref{fig:pro}b integrates both the carbon concentration and the MLE for the emitters’ orientations, as evident from these maps, incorporating the MLE results in sharper and more precise estimates of the vertical positions of the color centers.

\section{Geometry approximation of maximum strain}
\label{sec:gap}

\begin{figure*}[tbhp]
	\centering
	\includegraphics[width=\linewidth]{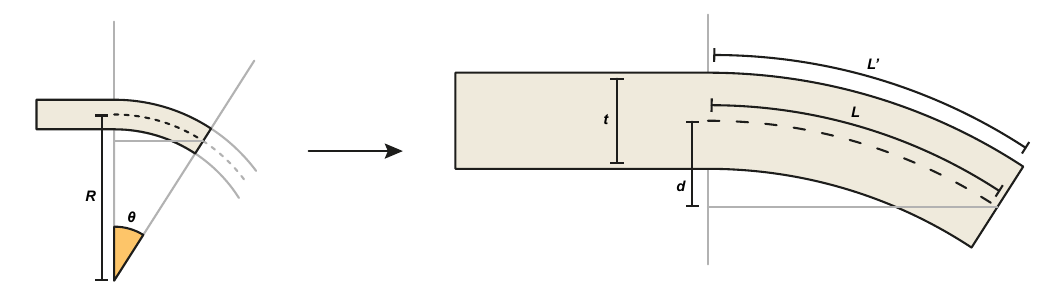}
    \caption{Lateral (XZ) cross-section schematic illustrating the geometrical parameters of the bent cantilever waveguide.}
    \label{fig:ssg}
\end{figure*}

To estimate the maximum strain achievable with the cantilever structure proposed in this work, the Euler–Bernoulli beam theory can be used under the assumption that the small-strain condition is valid, as commonly done in MEMS cantilever models \cite{senturia_structures_2001, obrien_mems_2001}. Alternatively, the same result can be obtained through a simple geometric argument. Referring to the structure depicted in Supplementary Fig.~\ref{fig:ssg}, \( R \) denotes the radius of curvature, \( \theta \) the angle subtended by the curvature of the cantilever, \( L \) the unstrained length of the cantilever, \( L' \) the strained length (at the top surface of the cantilever), \( d \) the vertical displacement, and \( t \) the thickness of the cantilever. The maximum strain on the upper side of the cantilever is given by
\[
\epsilon_\text{max} = \frac{\Delta L}{L} = \frac{L' - L}{L} = \frac{L'}{L} - 1.
\]
By substituting \( L' = (R + t/2)\,\theta \) and \( L = R\theta \), we find
\[
\epsilon_\text{max} = \frac{t}{2R},
\]
which corresponds to the result obtained with the Euler-Bernoulli theory. To express \( R \) as a function of displacement \( d \), we use \( d = R\,(1 - \cos\theta) \). For small \( \theta \), \( d \sim R\theta^2/2 \), yielding \( \theta = \sqrt{2d/R} \). With \( \theta = L/R \), we derive \( R = L^2/(2d) \). Thus, \( \epsilon_\text{max} \) becomes 
\[
\epsilon_\text{max} = \frac{td}{L^2}.
\]
Using \( d \) as the gap between the device layer and the silicon handle, we calculate a maximum strain of approximately \( 1.1~\text{m}\epsilon \). If \( d \) is taken as the pull-in displacement, equal to a third of the gap according to the parallel plate capacitor model \cite{senturia_structures_2001}, \( \epsilon_\text{max} \) reduces to \( 367~\mu\epsilon \), which closely aligns with finite-element-method simulations. 

\section{Poisson’s ratio in the laboratory coordinate system}
\label{sec:prl}

In Fig.~4c of the main text, the coordinate system is depicted, with the cantilever aligned along the \(y\)-direction in the laboratory reference frame. Stress is applied along the cantilever, corresponding to the crystal (\([1\bar{1}0]\)) direction with the resulting strain in the \(x\)- and \(z\)-directions scaled relative to the strain in the \(y\)-direction by Poisson’s ratio.

To compute Poisson’s ratio between the \(y\) and \(x\) directions (\([1\bar{1}0]\) and \([110]\)) and between the \(y\) and \(z\) directions (\([1\bar{1}0]\) and \([001]\)), we rotate the silicon elastic tensor to align it with the laboratory coordinate system. Specifically, we rotate the strain and stress directions by \(\theta = 45^\circ\) around the \(z\)-axis. The transformation is achieved by converting the stiffness matrix from Voigt notation (\(C_V\)) to the corresponding fourth-order tensor (\(C_T\)) with the Voigt-Tensor map. The rotation is then applied using the following expression in Einstein's notation

\[
C_{T,ijkl}' = R_{im} R_{jn} R_{ko} R_{lp} C_{T,mnop},
\]

where \(R\) is the 3×3 rotation matrix corresponding to the rotation around the $z$-axis, defined as

\[
R(\theta) = \begin{bmatrix}
\cos{\theta} & -\sin{\theta} & 0 \\
\sin{\theta} & \cos{\theta} & 0 \\
0 & 0 & 1
\end{bmatrix}.
\]

The rotated tensor \(C_T'\) is then transformed back to Voigt notation (\(C_V'\)) using the inverse Tensor-Voigt map.

Starting from the elasticity tensor in the crystal reference frame (in GPa)

\[
\begin{bmatrix}
165.7 & 63.9 & 63.9 & 0 & 0 & 0 \\
63.9 & 165.7 & 63.9 & 0 & 0 & 0 \\
63.9 & 63.9 & 165.7 & 0 & 0 & 0 \\
0 & 0 & 0 & 79.6 & 0 & 0 \\
0 & 0 & 0 & 0 & 79.6 & 0 \\
0 & 0 & 0 & 0 & 0 & 79.6
\end{bmatrix},
\]

we obtain the rotated stiffness matrix in the laboratory coordinate system (in GPa), as found in \cite{hopcroft_what_2010}

\[
\begin{bmatrix}
194.4 & 35.2 & 63.9 & 0 & 0 & 0 \\
35.2 & 194.4 & 63.9 & 0 & 0 & 0 \\
63.9 & 63.9 & 165.7 & 0 & 0 & 0 \\
0 & 0 & 0 & 79.6 & 0 & 0 \\
0 & 0 & 0 & 0 & 79.6 & 0 \\
0 & 0 & 0 & 0 & 0 & 50.9
\end{bmatrix}.
\]

From this, the Poisson’s ratios are computed as

\[
\nu_{y'x'} = -\frac{C_{12}}{C_{11}} = 0.0622 \qquad \nu_{y'z'} = -\frac{C_{13}}{C_{11}} = 0.3617.
\]

With Poisson’s ratios, we obtain the perpendicular strain components to compute the effect of those components on the emitter’s spectral shift using the piezospectroscopic model. The perpendicular strain components (in \(x\) and \(z\)) correspond to those found in the numerical simulations.

\section{Piezospectroscopic model}
\label{sec:sPO}
\subsection{Review piezospectroscopic model}
Given the symmetry of the G-center, the point group \(T_{d}\)  can be used to compute the rotational matrices required to transform the external strain tensor for each orientation. Starting from the identity orientation (E) referred to as defect plane (110), the symmetry operations consist of three \(C_{2}\) and eight \( C_3 \) rotations, which define the corresponding rotational axes and angles. The Rodrigues' rotation formula is employed to construct these rotational matrices. By applying the matrices to the external strain tensor, initially defined in crystal coordinates, the strain response for each rotated orientation can be determined. This approach ensures that all possible strain responses for different defect orientations are accounted for.
To validate this method and the piezospectroscipic model, we first reproduced the plots and fitting presented by Foy et al. \cite{foy_uniaxial_1981}, as shown in Supplementary Fig.~\ref{fig:sopF}.  As part of the standard procedure, uniaxial strain was applied along three principal axes directions: [$001$], [$110$], and [$111$], to an ensemble of G-centers in a Si cubic sample. The fitting of the data provided the extracted valued of \(A_p\) coefficients, presented in Foy's work, equal to (\(A_1\) = 13.4  \(A_2\) = -10.7, \(A_3\) = 4.8, \(A_4\) = \(\pm\)9.6)  \(\frac{\text{meV}}{\text{GPa}}\). Furthermore, the number of energy splittings observed for each direction provides additional confirmation of the monoclinic-I symmetry of the defect. The reproduced plots and energy splittings align with the original results, demonstrating the consistency of the methodology. The \(T_d\) symmetry rotations were applied to determine all possible orientations and corresponding strain transformations.

\begin{figure*}[tbhp]
	\centering
	\includegraphics[width=\linewidth]{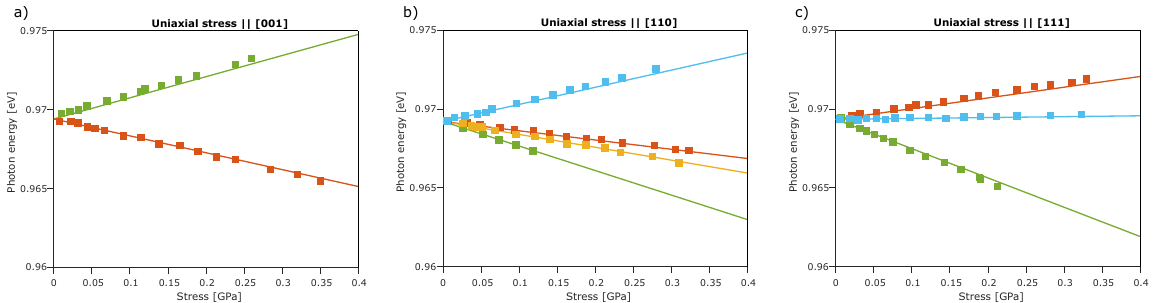}
    \caption{Experimental data from Foy et al. \cite{foy_uniaxial_1981} of uniaxial stress on G-center ensemble, piezospectroscopic model for fitting. \textbf{a)} Uniaxial stress on [$001$] direction, splitting into two sub-groups; \textbf{b)} Uniaxial stress on [$110$] direction, splitting into four sub-groups; \textbf{c)} Uniaxial stress on [$111$] direction, splitting into three sub-groups. The splitting represents the different stress responses for all the orientations. The initial stress is defined in crystal coordinates and then rotated for each orientation through the definition of rotational matrixes based on the \(T_d\) symmetry.}
    \label{fig:sopF}
\end{figure*}

\subsection{Piezospectroscopic shift with [$\overline{1}10$] uniaxial strain}
Once the model is validated and the rotational matrices are confirmed to be effective, the actual strain tensor must be determined. The waveguide under study is aligned along the [$\overline{1}10$] axis, where the application of a voltage induces a deflection, generating uniaxial strain of magnitude \(D\) in the same direction. The corresponding strain components, including the vertical direction, can be calculated using FEM simulations. Assuming a crystal coordinate system where \(x\) aligns with the [$100$] direction, \(y\) with [$010$], and \(z\) with [$001$], the strain tensor in crystal coordinates can be expressed as a function of \(D\). In this configuration, the nonzero strain components are \(\epsilon_{xx}\), \(\epsilon_{yy}\), and \(\epsilon_{xy}\), with an additional \(\epsilon_{zz}\) component due to the Poisson’s ratio of silicon. The \(A_p\) coefficients are referred to the application of an external stress. To convert the computed strain to a stress tensor, the elastic matrix of silicon can be applied

\[
\boldsymbol{\varepsilon}^{\mathrm{ext}}(D) =
\begin{pmatrix}
\varepsilon_{xx}^{\mathrm{ext}} \\
\varepsilon_{yy}^{\mathrm{ext}} \\
\varepsilon_{zz}^{\mathrm{ext}} \\
\varepsilon_{yz}^{\mathrm{ext}} \\
\varepsilon_{zx}^{\mathrm{ext}} \\
\varepsilon_{xy}^{\mathrm{ext}}
\end{pmatrix}
=
D \cdot
\begin{pmatrix}
\frac{1}{2} \\
\frac{1}{2} \\
-\nu_{x,y} \\
0 \\
0 \\
-\frac{1}{2}
\end{pmatrix}.
\]

By applying the piezospectroscopic model to the computed strain tensor, we observe a splitting of the emission wavelength into four distinct responses, each characterized by a unique coefficient (slope). These responses are associated with four different shift rates under strain (\(s_1, s_2, s_3,\)  and \(s_4)\) extracted from the analysis shown in Supplementary Fig.~\ref{fig:sp4}. Specifically, the shift rates are determined as follows: \(s_1 = 3.05 \times 10^{-3}\), \(s_2 = 2.03 \times 10^{-3}\), \(s_3 = -1.27 \times 10^{-3}\), and \(s_4 = 0.74 \times 10^{-3}\) \(\frac{\text{nm}}{\mu\varepsilon}\).

\begin{figure*}[tbhp]
	\centering
	\includegraphics[scale=0.35]{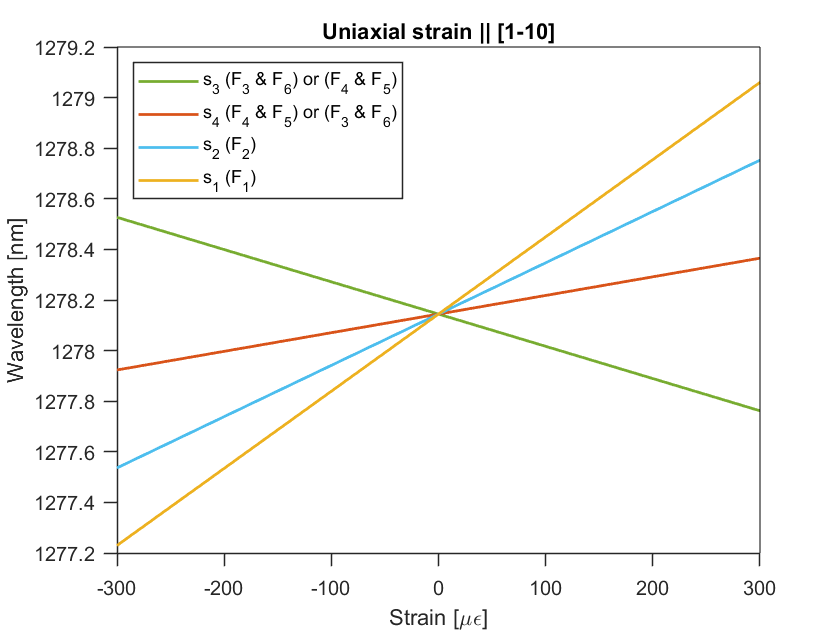}
    \caption{Strain response and energy splitting derived from the piezospectroscopic model, using the FEM-extracted strain tensor along the [$\overline{1}10$] direction. The plot shows four distinct splittings with their corresponding slopes (\(s_i\)) and rotational assignments. \(D\) is the applied strain that determines the strain components of the tensor.}
    \label{fig:sp4}
\end{figure*}

This analysis further identifies six distinct classes of orientations, summarized in Supplementary Fig.~\ref{fig:spT}, which group rotations resulting in equivalent final defect planes. By combining the rotational symmetry with the corresponding defect plane and strain response, we can assign a specific coefficient (shift rate of the emission wavelength as a function of uniaxial strain \(D\)) to each class.

\begin{figure*}[tbhp]
	\centering
	\includegraphics[width=\linewidth]{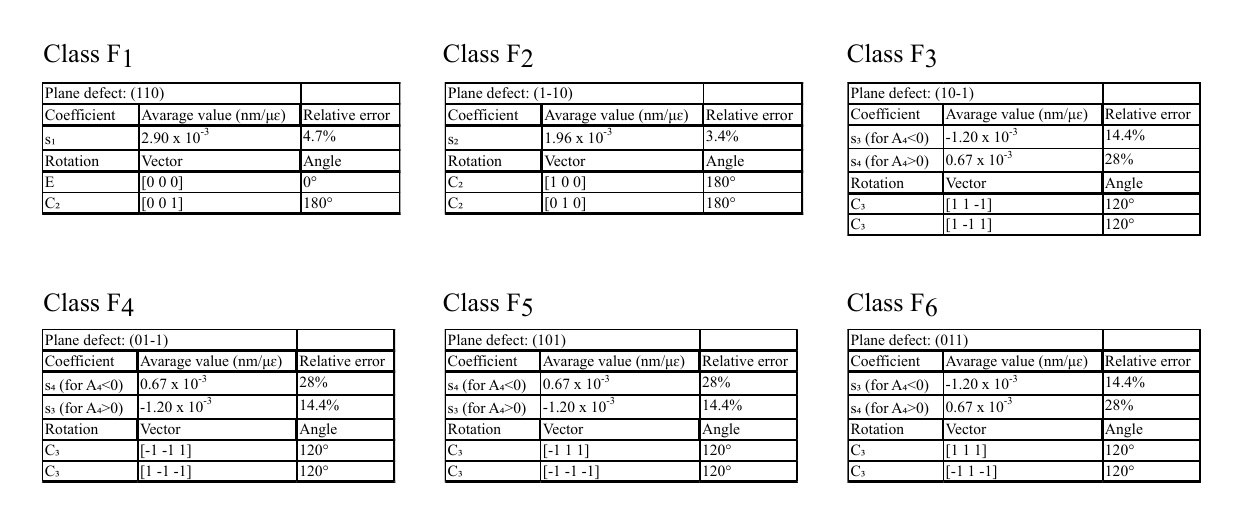}
    \caption{Summary of all possible symmetry operations, their corresponding orientations, and the resulting defect planes. Each sub-class, defined by its final defect plane, is associated with the respective shift rate variation (slope). The identity operation (\(E\)), representing the starting configuration, corresponds to the defect plane (110). The values of the \(s_i\) coefficients are presented, with the relative error reflecting variations in the \(A_p\) coefficients reported in early studies on the G-center. For each set of \(A_p\) coefficients, the \(s_i\) values were calculated, and the average values with their associated errors are reported.}
    \label{fig:spT}
\end{figure*}

Moreover, \(F_3\) and \(F_6\) have a slope \(s_3\) assuming the negative value of \(A_4\), or a slope \(s_4\) assuming positive \(A_4\). Vice versa for classes \(F_4\) and \(F_5\). This arises from the ambiguity in determining the sign of the coefficient \(A_4\).
\\During initial studies and discoveries of G-center, experimental investigations were performed on ensembles under uniaxial strain. These studies produced multiple sets of \(A_p\) coefficients \cite{foy_uniaxial_1981,davies_carbon-related_1983,thonke_new_1981}, indicating potential variations in the strain response. To address these discrepancies, the strain responses and corresponding \(s_i\) coefficients were calculated for all reported \(A_p\) sets in the literature. The average values and associated errors were then determined to account for potential variations and the corresponding error range. The final extracted coefficients are \(s_1 = 2.90 \pm 0.14  \times 10^{-3}\), \(s_2 = 1.96 \pm 0.07  \times 10^{-3}\), \(s_3 = -1.20 \pm 0.17  \times 10^{-3}\) and \(s_4 = 0.67 \pm 0.19  \times 10^{-3}\) \(\frac{\text{nm}}{\mu\varepsilon}\).

\subsection{T-center model and comparison}
This work is largely applicable, as it provides a framework for tuning other color centers in silicon. The T-center, another color center in silicon with the same point group of G-center (monoclinic \(C_{1h}\)), emits in the O-band, around $1326$~nm. Unlike the G-center, the T-center is characterized by two excited states, \(TX_0\) and \(TX_1\), where the strong asymmetry of the defect splits the TX state into two levels separated by $1.76$~meV.
\\A preliminary strain model for the T-center includes the same piezospectroscopic shift common to both excited states, with an additional term that models the defect potential as internal strain, responsible for this splitting \cite{clear_optical_2024}. By applying the strain tensor computed in this work to the T-center model, we can perform a theoretical comparison of the energy shifts. Due to the shared symmetry, the number of splittings remains the same as for the G-center, which is equal to four. However, the strain response is nonlinear to \(D\) due to the additional term in the Hamiltonian.
For the \(TX_0\) state, the maximum shift in wavelength, \(\Delta \lambda\) is \(\approx\) 1 nm, for the \(F_2\) class. In contrast, for the \(F_1\) class, characterized by the defect plane ($110$), the \(TX_1\) state exhibits a significantly larger strain response, shifting by almost $2$~nm, while the \(TX_0\) state shifts by \(\approx\) $0.4$~nm under a maximum applied strain of \(|D|=300\mu\varepsilon\). As for the G-center, the other classes show a relatively smaller strain response.

\end{document}